# Finely Tunable Thermal Expansion of NiTi by Stress-Induced Martensitic Transformation and Thermomechanical Training

Won Seok Choi [a,b,1], Won-Seok Ko [c,1,*], Yejun Park [a,1], Edward L. Pang [b], Jong-Hoon Park [d], Hye-Hyun Ahn [d], Yuji Ikeda [e], Pyuck-Pa Choi [a], and Blazej Grabowski [e]

[a] Department of Materials Science and Engineering, Korea Advanced Institute of Science and Technology, Daejeon 34141, Republic of Korea.
[b] Department of Materials Science and Engineering, Massachusetts Institute of Technology, 77 Massachusetts Avenue, Cambridge, MA 02139, USA.
[c] Department of Materials Science and Engineering, Korea University, Seoul 02841, Republic of Korea.
[d] Department of Materials Science and Engineering, Inha University, Incheon 22212, Republic of Korea.
[e] Institute for Materials Science, University of Stuttgart, Pfaffenwaldring 55, Stuttgart 70569, Germany.

**Abstract**

Tailoring the thermal expansion of martensitic materials by crystallographic texture and anisotropic variation of lattice parameters is a promising route to a flexible design of thermally stable systems. NiTi alloys are prototype materials in this respect, with shape-memory and superelastic properties owing to their thermoelastic martensitic transformations. Here, we propose a method to realize finely tunable coefficients of thermal expansion (CTE) for the NiTi alloy based upon a special combination of mechanical and thermal training. We achieve a near-zero in-plane CTE that is smaller in value than that of the FeNi-based Invar alloy. Atomistic simulations and theoretical calculations guide the method design and clarify the underlying mechanisms of the relationship between the processing conditions, the microstructural evolution, and the thermal expansion behavior. The directions for further, finer adjustments of the CTE without constraints on the shape of the materials are indicated.

[*] Corresponding author: wonsko@korea.ac.kr (Won-Seok Ko)
[1] These authors contributed equally to this manuscript.

# 1. Introduction

The thermal expansion of materials needs to be well-controlled in the design of precision instruments since they require high dimensional stability concerning temperature changes. In general, however, thermal expansion, *i.e.*, the expansion of materials upon heating due to quasiharmonic and anharmonic atomic vibrations [1], is challenging to control. Metals, for example, typically have a coefficient of linear thermal expansion (CTE) higher than ~10 × $10^{-6}$ $K^{-1}$ (*cf.*, Fig. 1). Representative exceptions are FeNi-based Invar alloys which exhibit a unique near-zero thermal expansion (ZTE), specifically a CTE of ~1.2 × $10^{-6}$ $K^{-1}$ between 297 K and 373 K [2, 3] (red dot in Fig. 1). These alloys have thus been the de facto standard for developing precision instruments [4-8]. Still, advances in controlling the CTE have also been achieved in other material classes. For example, materials with zero and negative thermal expansion (ZTE/NTE) were obtained based on flexible networks in rigid atomic structures of various ceramics [9-11] and using atomic-radius contraction in $Sm_{2.75}C_{60}$ [12].

A recently proposed concept for tailoring thermal expansion is based on crystallographic texture and anisotropic variation of lattice parameters of materials that exhibit martensitic phase transformations, such as shape-memory alloys (SMAs). Monroe *et al.* [13] demonstrated that crystallographic texture manipulation by macroscopic straining can be utilized to tailor thermal expansion. Significant changes in thermal expansion can be achieved in TiNb SMAs by deformation, *e.g.*, rolling [14]. Ahadi *et al.* [15] showed that severe plastic deformation develops deformation texture and induces anisotropy into the thermal expansion of NiTi SMAs. They explained that the coexistence of textured B2 austenite and B19′ martensite is the main reason for the anisotropic thermal expansion, and they obtained an almost zero thermal expansion (~-0.53 × $10^{-6}$ $K^{-1}$) for directions at angles of 30 – 45° with respect to the rolling direction. Overall, the available results [13-15] clarify the effect of mechanical training on the tailored thermal expansion of SMAs: the external stress induces either a transformation from austenite to martensite or a reorientation of existing martensitic variants. These variants are preferentially oriented depending on the direction of the applied stress, resulting in macroscopic texture. For example, Molnárová *et al.* [16] and Bian *et al.* [17] experimentally observed the development of fiber texture in NiTi wires under tensile stress using transmission electron microscopy (TEM) and in-situ synchrotron X-ray diffraction, respectively. Similarly, Chen *et al*. [18] demonstrated through molecular dynamics simulations that specific variants preferentially grow under compressive stress in nanocrystalline NiTi.

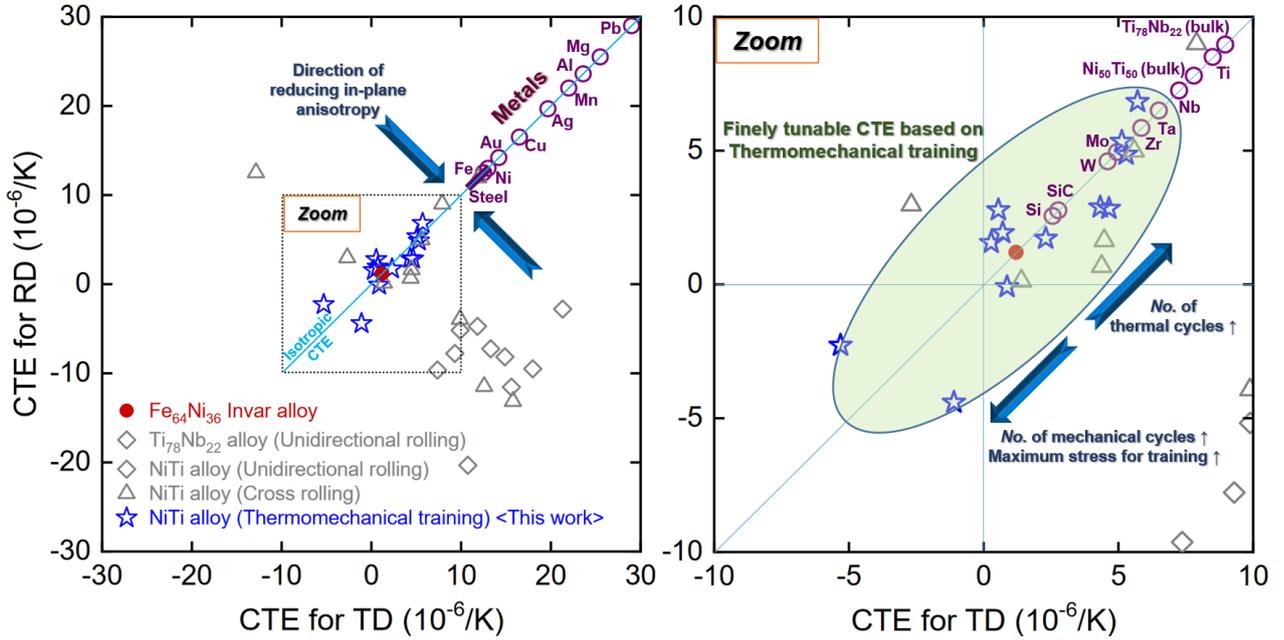

**Figure 1.** CTE map for various materials. The map expresses the anisotropy in the in-plane thermal expansion between the rolling direction (RD) and the transverse direction (TD) of martensitic materials prepared by unidirectional rolling [13, 15, 19] and cross-rolling [20, 21]. For the present values (stars), which correspond to thermomechanical training without a rolling direction, RD and TD reflect two arbitrary directions orthogonal to the compression axis. Traditional materials with an isotropic positive thermal expansion are located on the diagonal.

From an engineering perspective, producing materials in the form of plates, sheets, or thin films with ZTE along directions normal to the rolling or deposition direction (in-plane ZTE) and wires or rods with ZTE along the longitudinal direction is desirable. However, a critical limitation of current production methods is that the obtained thermal expansion is highly anisotropic in the plane normal to the rolling direction of sheets (*cf*., gray symbols in Fig. 1). For example, the work by Ahadi *et al*. [15] demonstrated that CTEs of severely cold-rolled sheets along the rolling and transverse directions are ~ -11 × $10^{-6}$ $K^{-1}$ and ~21 × $10^{-6}$ $K^{-1}$, respectively. Applications of such materials with strong in-plane anisotropy are limited compared to the FeNi-based Invar alloy with its volumetric ZTE. Recent works by Li *et al*. [20, 21] demonstrated that the anisotropy in the in-plane CTE can be reduced to a narrower range (rolling direction: 3.0 × $10^{-6}$ $K^{-1}$ and 0.13 × $10^{-6}$ $K^{-1}$; transverse direction: -2.7 × $10^{-6}$ $K^{-1}$ and 1.4 × $10^{-6}$ $K^{-1}$) by applying multiple cross-rolling processes. However, the obtained CTE values are still larger in magnitude and distributed into a broader range of positive and negative values compared to the FeNi-based Invar alloy (~1.2 × 10$^{-6}$ K$^{-1}$). Moreover, severe deformation can limit the application of the materials to bulk structural properties since cold workability constrains the maximum amount of thickness reduction. In many applications, such as electronic packaging [22-24], shadow masking of displays [25-27], and MEMs/NEMs [28-30], severe cold-rolling cannot be thus applied.

The present study introduces a simple but effective approach to freely adjust the CTE of SMAs over a wide range while significantly reducing the in-plane anisotropy. Focusing on the most widely used SMA, *i.e.*, the NiTi alloy, as a prototype example, we demonstrate that an adjustable CTE can be realized by applying an optimized thermomechanical training procedure, combining cyclic mechanical and thermal loading *without* severe plastic deformation. Our approach is inspired by the two-way shape-memory effect, which is the ability of SMA materials to remember not only the macroscopic shape at high temperatures (where the austenite phase is stable) but also the shape at low temperatures (where martensite is stable). We provide an atomistic understanding of the relation between the processing conditions, the microstructural evolution, and the observed thermal expansion behavior by theoretical investigations, specifically molecular dynamics (MD) simulations and calculations based on the geometrically nonlinear theory of martensite (GNLTM) [31-33]. Finally, we experimentally demonstrate the effect of the derived processing conditions on the CTE of NiTi (star symbols in Fig. 1).

## 2. Methodology

### 2.1. Molecular dynamics (MD) simulations

The MD simulations were performed with a second nearest neighbor modified embedded-atom method (2NN MEAM) potential for the Ni-Ti binary system [34]. The radial cutoff distance was 5.0 Å, which is larger than the second nearest-neighbor distance of the B2 NiTi structure. The here used interatomic potential was developed in the previous study [34] with the aim of reproducing the temperature- and stress-induced phase transformations in NiTi alloys with equiatomic composition. The potential accurately reproduces the diffusionless phase transformation between the B2 (cubic) austenite and B19′ (monoclinic) martensite phase, as well as the fundamental physical properties (structural, thermodynamic, and defect properties) of the intermetallic compounds and solid solutions. The reliability of this interatomic potential in reproducing the SMA properties has been repeatedly verified in the literature, *e.g.*, for nanocrystalline SMAs [39], SMA nanoprecipitates embedded in a stiff non-transforming matrix[55], freestanding SMA nanoparticles [56], SMA nanowires [57], SMA nanopillars [58, 59], and SMAs with complex microstructures [60]. We further examined the reliability of the potential by focusing on properties related to the thermal expansion, as shown in Note S1, Supplementary materials. A detailed formulation of the 2NN MEAM formalism is available in the literature [61].

Nanocrystalline cells were prepared using the Voronoi construction method [62] with random crystallographic orientations and positions of seeds for each grain using the Atomsk [63] software. Initially, cube-shaped cells with 30 grains were generated based on the equiatomic ($Ni_{50}Ti_{50}$) B2 austenite structure. Table S1 summarizes the average grain diameter, number of grains, cell dimensions, and number of atoms of each generated cell.

The MD simulations were performed using the LAMMPS code [64] with a time step of 2 fs. Periodic boundary conditions were applied along all three dimensions, and the Nosé-Hoover thermostat and barostat [65, 66] were used for controlling temperature and pressure, respectively. During the MD runs, individual atomic positions, dimensions, and angles of the simulation cells were allowed to fully relax under the designated stress state. The generated cells were initially subjected to an energy minimization process based on the conjugate gradient method. Annealing at high temperatures was then applied to recover the initial B2 austenite structure. The thermal loading of the cell was induced by MD simulations in an isobaric-isothermal (*NPT*) ensemble at zero pressure. From the designated initial temperature, the temperature was gradually changed with cooling and heating rates of ±0.5 K/ps. During the MD runs, the volumes and dimensions of the cells at the specific temperature and their changes with temperature were recorded to observe the occurrence of phase transformations and to analyze the thermal expansion behavior. Before the mechanical loading, transformation temperatures for each nanocrystalline cell at zero stress were measured from MD runs of the temperature-induced phase transformation, as shown in Fig. S1. The mechanical loading of the cell was induced by MD simulations in an *NPT* ensemble under controlled pressure conditions. A stress-controlled uniaxial loading was applied by adjusting the pressure along one direction of the cells with a stress rate of 12.5 MPa/ps. The stress along this direction was gradually increased to the maximum value and decreased to zero, allowing full relaxation of individual atomic positions under a given stress state. During the loading, the pressure in the directions orthogonal to the loading direction was set to zero, and the cell angle between those directions was allowed to relax, leading to a dynamic response to the mechanical loading.

The microstructural evolution was visualized during the MD runs using the Polyhedral Template Matching (PTM) method [35] as implemented in the OVITO program [67]. Although the PTM algorithm was initially not developed to distinguish atoms corresponding to the B2 austenite and B19′ martensite structures, we confirmed that it is well suited to distinguish the martensite phase from the austenite phase and to identify the occurrence of the phase transformation. In the PTM pattern, atoms depicted in blue represent the B2 austenite structure. The B19′ martensite structure is mostly represented by red atoms. Undetermined regions, *i.e.*, surfaces and amorphous regions, are shown in gray, while some atoms in the crystalline regions, also represented by gray color, are present due to the thermal noise at finite temperatures. To emphasize the evolution of martensite variants, the local lattice orientation (LLO) of the crystalline regions was visualized based on the PTM algorithm [35]. The von Mises local shear invariant [43] of each atom was additionally calculated and visualized using the OVITO program to emphasize the local plastic deformation.

## 2.2. CTE calculations based on the GNLPTM

Martensite habit plane variants (HPVs) were computed using the geometrically nonlinear theory of martensite (GNLTM) [32, 33]. The following lattice parameters [31] were used for the B2 austenite and B19′ martensite phases, respectively: $a_{B2}$ = 3.015 Å; $a_{B19'}$ = 2.889 Å, $b_{B19'}$ = 4.120 Å, $c_{B19'}$ = 4.622 Å, and $\beta_{B19'}$ = 96.8°. In the case of untwinned martensite, the stretch tensor gives the total deformation in the calculations. In the case of twinned martensite, the total deformation is an average over two twin-related correspondence variants coupled with rigid body rotations, as given by the habit plane equation [32, 33].

Variant selection was performed using the maximum work criterion [68]. CTE values for a given martensite correspondence variant in the loading direction were computed as:

$$CTE = \hat{l}^T \alpha \hat{l},  \qquad (1)$$

where $\hat{l}$ is a unit vector in the loading direction (in the reference frame of the CTE tensor) and the CTE tensor $\alpha$ (in $10^{-6}$ K$^{-1}$) is given by[69]:

$$\alpha = \begin{bmatrix} 14.9 & 0 & 36.4 \\ 0 & 43.8 & 0 \\ 36.4 & 0 & -39.4 \end{bmatrix}. \qquad (2)$$

Note that these values differ from those given in the reference, as we transformed the coordinate system to match the typical convention [70]. In the case of twinned martensite, the CTE is taken as the weighted average according to the calculated twin fractions of each correspondence variant.

## 2.3. Experimental validation

A nearly equiatomic NiTi alloy was fabricated by vacuum arc re-melting (VAR) from Ni and Ti of 99.99% purity with 3 × 3 mm pellet shape (iNexus Inc.). Ingots were re-melted five times and homogenized in Ar atmosphere at 900 °C for 24 h and water quenched. The 6 × 6 × 6 mm³ cubes were machined by electrical discharge machining (EDM), and the surface oxide was mechanically milled (Fig. 6f). Differential scanning calorimetry (DSC) was conducted to determine transformation temperatures in a METZSCH DSC 214 Polyma instrument under flowing nitrogen (30 ml/min) at a heating and cooling rate of 10 °C/min. The martensite phase was identified by X-ray diffraction (XRD) (SmartLab, RIGAKU) with monochromatic Cu-Kα1 radiation (45 kV) in the 30–100° range. Martensite microstructures were characterized using transmission electron microscopy (TEM, JEOL 2100F operated at 200 kV). TEM samples were extracted by focused ion beam (FIB) milling (FEI Helios NanoLab 600i instrument), and final cleaning was conducted at 5 keV. The coefficient of thermal expansion (CTE) was measured in a thermomechanical analyzer (NETZSCH, TMA 402 F1) between -40 °C and 40 °C. The samples were held at -50 °C for 10 minutes to equilibrate the sample temperature, followed by slow heating at 3 °C/min; the absence of thermal lag was confirmed by the excellent agreement of the austenite start temperature (92.8 °C) with that measured using DSC. Thermal cycling after training was also carried out in the TMA at a 3 °C/min rate. Strains were computed from the crosshead displacement.

## 3. Results and Discussion

### 3.1. Theoretical Analysis of the Thermal Expansion Anisotropy of NiTi SMAs

We examine with MD simulations the atomistic origin of the experimentally confirmed (Section 3.3) unique thermal expansion behavior of NiTi SMAs. A series of MD runs based on an established interatomic potential for the Ni-Ti binary system [34] (validation of the potential is given in Note S1, Supplementary materials) is utilized to provide a detailed atomistic understanding of the correlation between the processing conditions, the microstructural evolution, and the thermal expansion behavior.

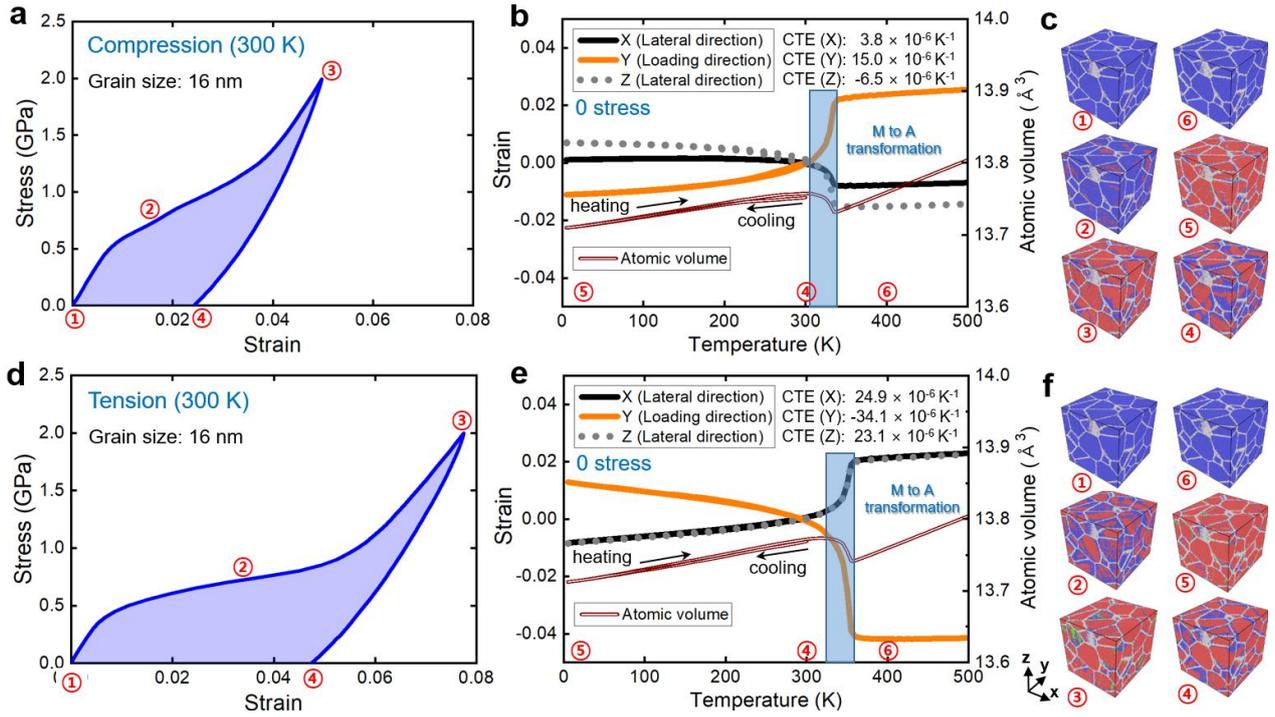

**Figure 2.** Impact of compressive and tensile loading on the thermal expansion behavior, predicted by the present MD simulations. (a,d) Stress-strain responses of the nanocrystalline cell with an average grain size of 16 nm under compressive and tensile loading and unloading at 300 K. (b,e) Temperature dependence of the directional strain and atomic volume of the nanocrystalline cell during the subsequent thermal loading (300 → 5 → 500 K) at zero external stress. The reference of the strain values is the cell at 300 K. The CTE measured in the temperature range of 5 – 100 K is indicated for each direction. (c,f) Corresponding atomic configurations visualized by the PTM algorithm [35]. In each snapshot, blue atoms correspond to the B2 austenite structure, red atoms to the B19′ martensite structure, and gray atoms to the grain and domain boundaries.

First, we analyze the effect of mechanical strain in a form that is easy to apply in manufacturing (*e.g.*, uniaxial compression or tension). We utilize nanocrystalline cells of equiatomic NiTi ($Ni_{50}Ti_{50}$) containing 30 randomly oriented grains, initially in the B2 austenite phase (Table S1). Simulation cells with different average grain diameters are stressed in compression or tension to induce the martensitic transformation. Figs. 2a and 2d show the resulting stress-strain response of a nanocrystalline cell with an average grain size of 16 nm at 300 K under compressive and tensile loading, respectively. The temperature of 300 K is slightly higher than the austenite finish temperature ($A_f$) of the selected nanocrystalline cell (Fig. S1). A plateau-like deviation from linear elasticity can be seen in the stress-strain curves, indicative of a stress-induced transformation from the initial B2 austenite phase to B19′ martensite. Because of the nanocrystalline grain size, the plateau is inclined rather than flat, as often reported in large-grained SMAs. The present findings are consistent with previous experimental [36-38] and MD [39] studies on nanocrystalline NiTi SMAs. The maximum reachable strain corresponding to the maximum stress of 2 GPa is higher for tensile loading (~0.076) than for compressive loading (~0.050). This tension-compression asymmetry is consistent with experimental results for NiTi SMAs [40, 41].

Figs. 2c and 2f show atomistic configurations of the nanocrystalline cell during the deformation process: in the initial state (①), during compressive/tensile loading (②), at the maximum compressive/tensile loading (③), and at the fully unloaded state (④). The formation of the martensite phase under mechanical loading is incomplete, and a substantial amount of *retained austenite* (blue color) is observed near the grain boundary region at the maximum loading condition. The amount of retained austenite depends on the orientation of the grains with respect to the loading condition [38, 39]. After complete unloading, the resulting stress-strain responses exhibit a significant amount of residual strain due to the presence of *retained martensite*. The retained martensite is a consequence of the utilized temperature, which is only slightly higher than the $A_f$ temperature of the selected nanocrystalline cell. Additional tests confirm that no severe plastic deformation via dislocation slip is involved.

Starting with the post-deformation microstructures containing the retained martensite phase, the thermal expansion behavior of the nanocrystalline cells is investigated by applying a subsequent thermal loading.

Figs. 2b and 2e show the temperature dependence of the atomic volume (brown curves) during cooling (300 K to 5 K) and reheating (5 K to 500 K) of the pre-deformed nanocrystalline cell with an average grain size of 16 nm. During cooling, the martensite phase grows and consumes the austenite phase (⑤ in Fig. 2c, f) while maintaining the same twin and domain structure in each grain, as has been induced by the precedent mechanical loading (details on the microstructural evolution are given in Note S2, Supplementary materials). The dominant twinning mode observed in the present MD simulations is based on (001)B19′ compound twins, with the herringbone-like arrangement of these twins depending on the size of the grains. When the material is reheated, the martensite phase shrinks to form the austenite phase, and the reverse B19′ → B2 transformation is completed (⑥ in Fig. 2c, f), indicating a negative volume change consistent with previous experiments [42].

An important insight from the thermal loading simulations is the anisotropy in the thermal expansion induced by the different loading conditions. Figs. 2b and 2e detail the temperature dependence of the strain for each direction, taking the dimension of the cell at 300 K obtained after the mechanical loading as a reference. For the nanocrystalline cell prepared with compressive loading (Fig. 2b), the direction along the compression axis ($Y$) exhibits a positive thermal expansion. One of the directions perpendicular to the compressive axis ($Z$) exhibits a negative thermal expansion. The other perpendicular direction ($X$) shows a nearly constant thermal expansion at lower temperatures, and it gradually changes to a negative thermal expansion when the temperature approaches the $A_f$ temperature. This trend is consistent with previous experimental results on nanocrystalline NiTi prepared by severe cold-rolling [15]. Focusing now on the thermal expansion of the nanocrystalline cell prepared with tensile loading, a very different anisotropy is observed. As shown in Fig. 2e, the tensile loading direction ($Y$) exhibits a negative thermal expansion, whereas both perpendicular directions ($X$ and $Z$) exhibit a positive thermal expansion. The MD runs are based on simulation cells with a small number of grains (30) compared to typical experiments on bulk polycrystalline specimens, so the observed trends must be statistically verified. We have, therefore, additionally performed independent MD runs using different directions of initial mechanical loadings. The same tendencies (*i.e.*, NTE in one direction perpendicular to the compression axis and NTE in the tensile direction) are consistently found for different loading directions (Fig. S5).

It is important to note that the CTE of the nanocrystalline cells can be affected by continuous phase transformations during the thermal loading process. The volume expansion exhibits a slim hysteresis during the cooling and reheating processes (*i.e.*, a mismatch between the corresponding curves in Figs. 2b and 2e). This indicates a continuous formation (disappearance) of B19′ martensite in cooling (reheating), as observed and explained by previous experiments for a nanocrystalline NiTi alloy[20]. The phase fractions derived from the atomistic snapshots verify a continuous evolution of the martensite phase (Fig. S6). Even though such a continuous phase transformation can alter the CTE of the NiTi alloy, it mostly affects the CTE in an isotropic manner (*i.e.*, volume change). The anisotropy of the CTE is mainly associated with the twin/domain structure of each grain as induced by the preceding mechanical loading. We have additionally performed independent MD runs using nanocrystalline cells with different average grain sizes (8 – 14 nm). Although various degrees of the continuous phase transformation are found due to the different grain boundary region-to-volume ratios, the same tendencies of the anisotropic CTE are consistently present for every nanocrystalline cell regardless of the grain size (see Figs. S7 and S8).

Another important consideration is required to clarify the potential impact of the twinning mode on the deformation-induced anisotropy in the CTE. The nanocrystalline cells investigated in the MD simulations promote the (001) compound twinning in a herringbone microstructure. In polycrystalline materials with much larger grain sizes, other twinning modes are possible, with the Type II twinning mode being particularly relevant. Since MD simulations of such large grain sizes are impossible, we resort here to the phenomenological geometrically nonlinear theory of martensite (GNLPTM). Specifically, we examine the effect of mechanical loading on the CTE for different twinning modes that can occur in the NiTi alloy depending on the grain size (Note S4, Supplementary materials). The GNLPTM results show that the twinning mode affects the CTE only quantitatively. The qualitative asymmetry trends predicted by the MD simulations do not depend on the type of twinning mode, *i.e.*, there is always PTE for the compressive direction and NTE for the tensile direction (Fig. S12).

Overall, the presented theoretical investigations demonstrate that the thermal expansion of NiTi alloys can be tuned *without* severe plastic deformation by moderate mechanical loads selected according to the practical use of the materials. For example, if the material is produced in the form of sheets, plates, or thin films, compressive loading induces NTE/ZTE along the in-plane directions. In contrast, tensile loading induces NTE/ZTE along the longitudinal direction of materials in the form of wires or rods. Section 3.3 will present a practical experimental procedure based on these insights.

## 3.2. Finely Tunable Thermal Expansion by Cyclic Mechanical and Thermal Training

We have demonstrated that the thermal expansion of NiTi alloys can be tailored without severe plastic deformation. This is beneficial because severe plastic deformation would trigger a significant in-plane anisotropy. According to the MD results, a single mechanical loading and unloading process is enough to provide different ZTE/PTE combinations. Still, a single adjustable processing condition (*i.e.*, only the amount of mechanical loading) is too restrictive to achieve all desired CTE properties flexibly. In this section, we therefore present the impact of additional processing conditions on the thermal expansion (*i.e.*, cyclic mechanical and thermal loading). We examine why such processing conditions lead to a diverse and finely tunable thermal expansion behavior.

Figs. 3 and 4 show the results of MD simulations for cyclic mechanical compression and tension, respectively. The stress-strain response for five cyclic compressive and tensile loadings at 300 K is presented in Figs. 3a and 4a, respectively. Figs. 3b and 4b show the temperature dependence of the atomic volume and thermal strain for each direction during the following thermal loading process. The sequence of heating and cooling is reversed with respect to the thermal process discussed in the previous section (Sec. 3.1), *i.e.*, the heating of the nanocrystalline cells to high temperature (500 K) is conducted first (③→④→⑤ in Figs. 3b and 4b). The retained martensite phase obtained from the mechanical loading completely transforms back into the B2 austenite phase. During the subsequent cooling process (⑤→⑥ in Figs. 3b and 4b), the austenite to martensite transformation is accompanied by a discontinuous jump in the volume.

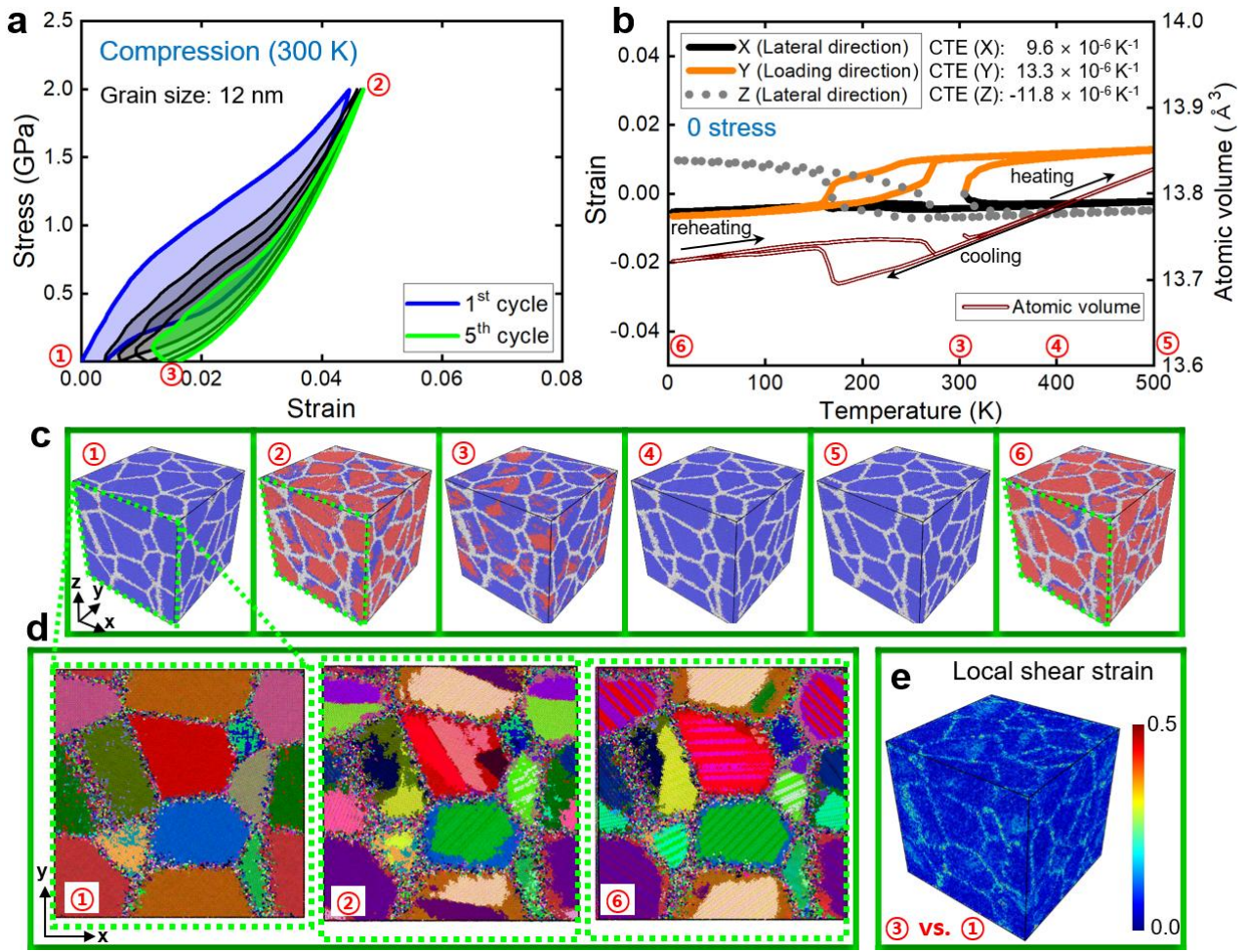

**Figure 3.** Impact of cyclic compression on the thermal expansion behavior, predicted by the present MD simulations. (a) Stress-strain responses of the nanocrystalline cell with an average grain size of 12 nm under 5 cyclic mechanical (compressive) loadings at 300 K. (b) Temperature dependence of the directional strain and atomic volume of the nanocrystalline cell during the subsequent thermal loading (300 → 500 → 5 → 500 K) at zero external stress. The reference of the strain values is the cell at 300 K. The coefficient of linear thermal expansion (CTE) measured in the temperature range of 5 – 100 K is indicated for each direction. (c) Corresponding atomic configurations visualized by the PTM algorithm [35]. In each snapshot, blue atoms correspond to the B2 austenite structure, red atoms to the B19′ martensite structure, and gray atoms to the grain and domain boundaries. (d) Evolution of microstructure during the mechanical and thermal loadings, identified by the local lattice orientations (LLO) of crystalline regions based on the PTM algorithm [35]. The color of atoms corresponds to specific crystal orientations. (e) The local shear strain [43] accumulated during the cyclic mechanical loading is visualized by comparing the initial (①) and final (③) configurations of the mechanical loading.

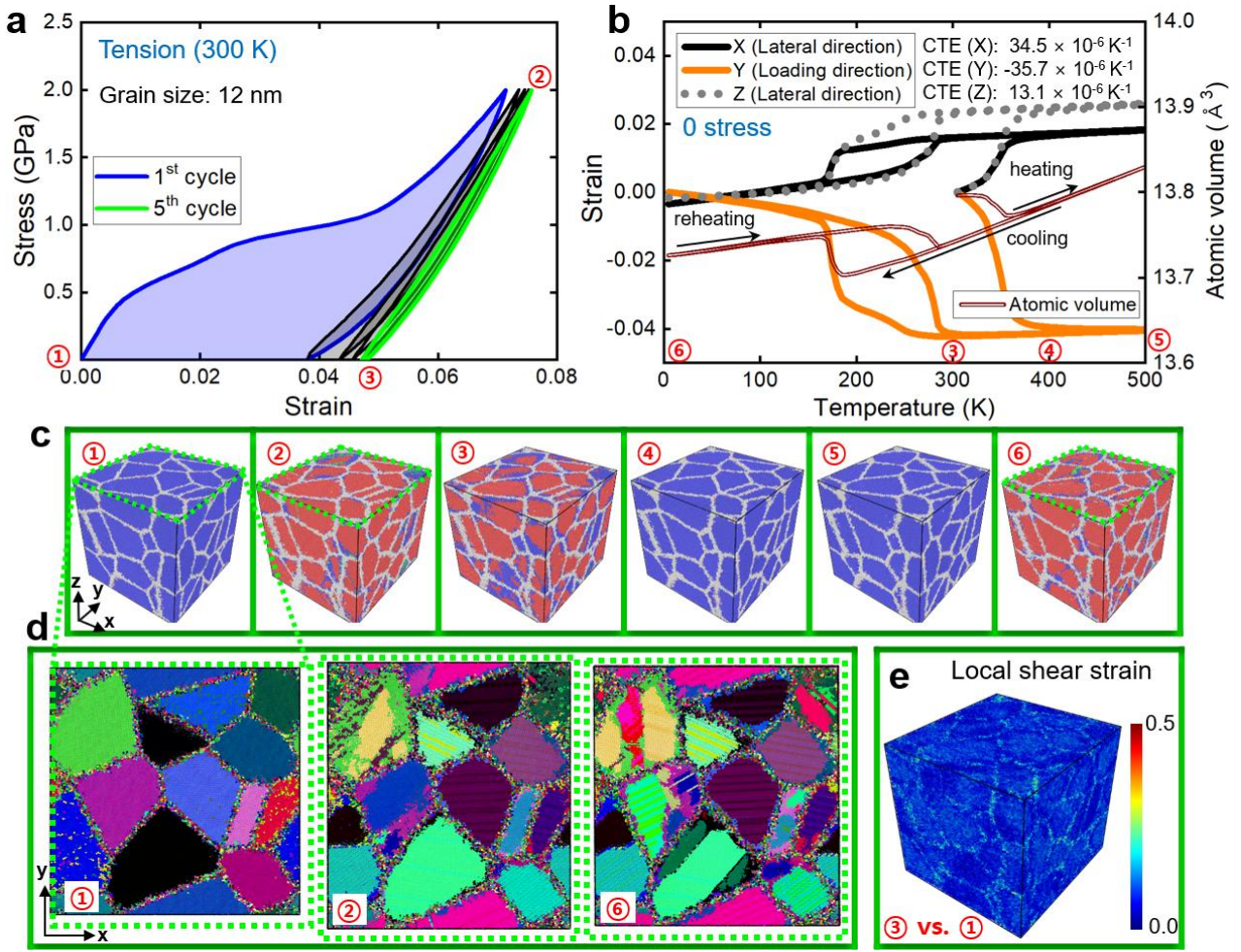

**Figure 4.** Impact of cyclic tension on the thermal expansion behavior, predicted by the present MD simulations. (a) Stress-strain responses of the nanocrystalline cell with an average grain size of 12 nm under 5 cyclic mechanical (tensile) loadings at 300 K. (b) Temperature dependence of the directional strain and atomic volume of the nanocrystalline cell during the subsequent thermal loading (300 → 500 → 5 → 500 K) at zero external stress. The reference of the strain values is the cell at 300 K. The coefficient of linear thermal expansion (CTE) measured in the temperature range of 5 – 100 K is indicated for each direction. (c) Corresponding atomic configurations visualized by the PTM algorithm [35]. In each snapshot, blue atoms correspond to the B2 austenite structure, red atoms to the B19′ martensite structure, and gray atoms to the grain and domain boundaries. (d) Evolution of the microstructure during the mechanical and thermal loading, identified by the local lattice orientation (LLO) of the crystalline regions based on the PTM algorithm [35]. The color of atoms corresponds to specific crystal orientations. (e) The local shear strain [43] accumulated during the cyclic mechanical loading is visualized by comparing the initial (①) and final (③) configurations of the mechanical loading.

A crucial insight from this thermal loading process is that the anisotropy in the thermal expansion is still present at temperatures below the austenite to martensite phase transformation, even though the cooling process starts from grains fully covered with the austenite phase (*cf.*, Figs. 3c and 4c). Notably, the preferred direction for the NTE is the same as for the nanocrystalline cells initially containing retained martensite (*cf.*, Fig. 2). For example, one of the directions perpendicular to the compressive loading axis (*Z*) exhibits a negative thermal expansion (Fig. 3b). Similarly, the tensile direction (*Y*) exhibits a negative thermal expansion (Fig. 4b). As shown by the local lattice orientation (LLO) analysis (Figs. 3d and 4d), the orientation patterns (indicated by the colors) of several grains at the maximum compressive stress of $5^{th}$ cycle (②) are identical to the orientation patterns after the subsequent heating and cooling processes (⑥). This result implies that the orientation and twin structure of the martensite phase are memorized during the initial mechanically induced transformation and then manifested in the subsequent thermally induced transformation. This behavior is analogous to the two-way shape-memory effect, which is the ability of SMA materials to remember not only the macroscopic shape at high temperatures but also at low temperatures, *i.e.*, below the martensitic transformation temperature.

As shown in Figs. 3a and 4a, the stress-strain response for the first cyclic loading exhibits a significant amount of residual strain at the fully unloaded state, primarily due to the presence of retained martensite. The residual strain gradually accumulates with an increasing number of cycles. The accumulation of residual strain is caused by plastic deformation of the amorphous-like grain boundary region, as revealed by the accumulated atomic local shear strain after the cyclic loading. The local shear strain shown in Figs. 3e and 4e verifies that the plastic deformation is localized at the grain boundaries. The plastic deformation accumulated during the cyclic mechanical loading serves to memorize the orientation of the nearby martensite phase with respect to its twin and domain structure induced by the mechanical loading. During the subsequent thermal loading process, martensite formation is guided by the accumulated plastic deformation into a specific twin and domain microstructure.

The MD results thus indicate that thermomechanical training can be an additional processing condition to achieve the desired thermal expansion behavior. To substantiate this finding, we show how the processing variables can be further customized. We perform additional MD simulations for different maximum mechanical loading and number of mechanical as well as thermal cycles. We focus on tensile loading, for which the statistical variance in the CTE results is smaller and, thus, a statistically significant tendency in the CTE dependence is computationally more straightforward to achieve.

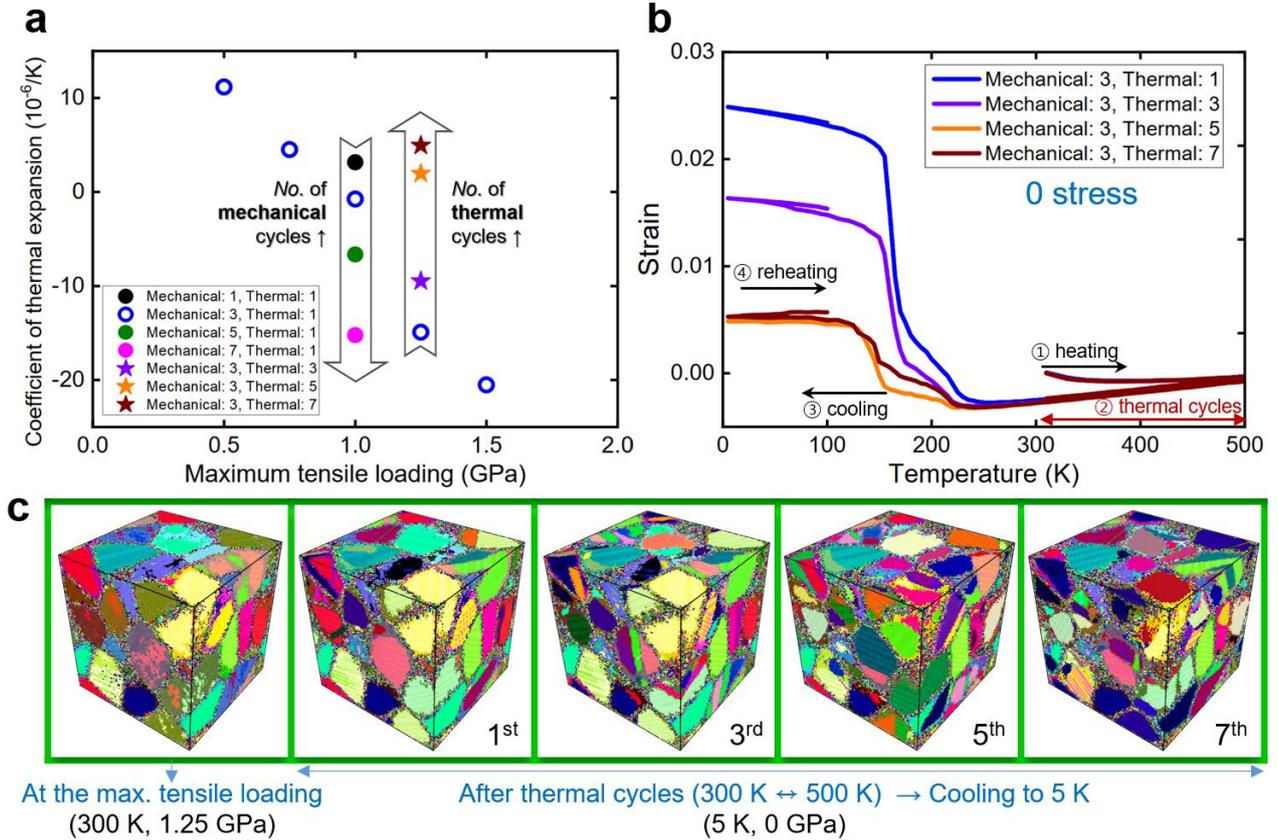

**Figure 5.** Impact of different thermomechanical training conditions on the thermal expansion behavior, predicted by the present MD simulations. (a) Variation of the CTE (5 – 100K) induced by the thermomechanical training of the nanocrystalline cell with an average grain size of 12 nm. Effects of different levels of the maximum tensile loading (0.50, 0.75, 1.00, 1.25, and 1.50 GPa), numbers of mechanical cycles (1, 3, 5, and 7) for the training, and numbers of thermal cycles (1, 3, 5, and 7) on the CTE values are presented. (b) Temperature dependence of the directional strain in the nanocrystalline cell during the thermal loading at zero external stress. (c) Evolution of the microstructure induced by the thermomechanical training, identified by the local lattice orientation (LLO) of the crystalline regions based on the PTM algorithm [35]. The color of atoms corresponds to specific crystal orientations.

Fig. 5 illustrates the impact of the processing variables on the thermal expansion behavior. The NTE along the tensile direction is more pronounced for higher maximum stress and more mechanical cycles. This result indicates that the twin and domain structures are progressively adapted into a form more suitable for manifesting the NTE upon increasing mechanical loading. In contrast, increasing the number of thermal cycles reduces the NTE along the tensile direction. As shown in Fig. 5c, the twin and domain structures adapted from the preceding mechanical cycles are gradually relieved and changed to a more random orientation distribution (*i.e.*, the colors of the grains increasingly deviate from the initial colors representing the microstructure after the maximum loading). This behavior is correlated with a decreasing NTE. The results show that the CTE of NiTi SMAs can be finely tuned by applying a specific thermomechanical training process and adjusting the relevant variables.

### 3.3. Experimental Demonstration of Tailoring the Thermal Expansion

Inspired by the theoretical results, we devised an experimental procedure to fine-tune the CTE of SMAs. We demonstrate the effectiveness of the processing protocol for the equiatomic NiTi SMA. The equiatomic composition ($Ni_{50}Ti_{50}$) exhibits relatively high transformation temperatures and maintains the martensite phase at ambient conditions. The as-homogenized samples exhibit transformation temperatures of $M_s$ = 340 K, $M_f$ = 315 K, $A_s$ = 366 K, and $A_f$ = 384 K (corresponding to the martensite start, martensite finish, austenite start, and austenite finish temperature, respectively; *cf.*, Figs. 6b). The differential scanning calorimetry (DSC) results in Fig. 6b indicate that the martensitic transformation occurs without the appearance of an intermediate *R* phase. At room temperature, the samples consist of entirely lath-shaped B19′ martensite (Fig. 6g) with prevalent <011> Type II twins (Figs. 6e and 6f), in agreement with the reported transformation-mediated twinning mode for solutionized NiTi alloys [44].

Bulk NiTi machined into 6 × 6 × 6 mm³ cubes is used to test the thermal expansion behavior along the three axes (*X*, *Y*, and *Z*) after thermomechanical training (Fig. 6b). The thermomechanical process (Fig. 6d) designed according to our theoretical predictions is as follows: the sample is 'trained' by heating to 393 K, slightly higher than its $A_f$ temperature (384 K), and compressing it at different levels of the maximum strain and for different numbers of cycles. We test three mechanical training conditions: (i) A target strain of 5% with three cycles, (ii) a target strain of 5% with ten cycles, and (iii) a target strain of 10% applied once without additional cycling. Hereafter, we call the samples 5%-M3, 5%-M10, and 10%-M1, respectively. The engineering stress-strain curves for the training are shown in Fig. S11. The material does not show significant superelastic behavior but rather training-induced residual strain. The accumulated residual strain during training is 0.9%, 3.4%, and 3.9% for 5%-M3, 5%-M10, and 10%-M1, respectively (Fig. S11). The calculated strain values with respect to the dimension of the as-trained samples are $\varepsilon_{5\%-M3}$ = 3.1%, $\varepsilon_{5\%-M10}$ = 7.7%, and $\varepsilon_{10\%-M1}$ = 7.7%, respectively.

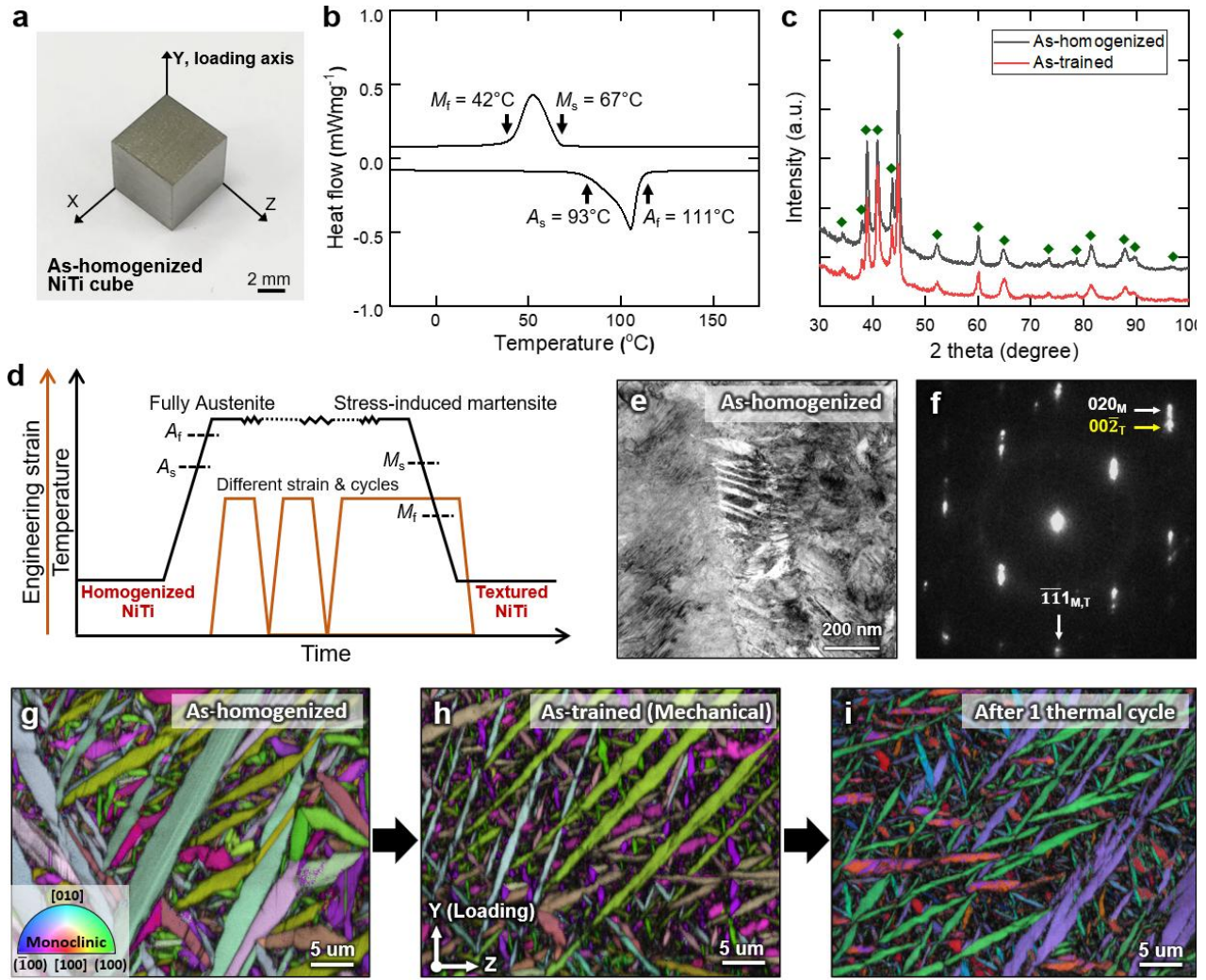

**Figure 6.** Sample information and analyzed results. (a) The as-homogenized 6 × 6 × 6 mm³ NiTi cube sample. (b) Phase transformation temperatures measured by DSC. (c) XRD results of the as-homogenized and mechanically trained materials (5%-M10). (d) A schematic of the thermomechanical process was designed based on considering the measured transformation temperatures. The sample was mechanically loaded above $A_f$ temperature and was cooled down with applied stress to stabilize stress-induced martensite. (e) TEM image and (f) diffraction pattern of the as-homogenized sample with a prevalent transformation twinning mode of <011> Type II. (g–i) Inverse pole figure maps from EBSD analysis of samples with fully lath-shaped B19′ martensite after the homogenization, mechanical training (5%-M10), and a single thermal cycle, respectively.

After training, the samples are cooled down while maintaining the applied stress to obtain a mechanically induced martensitic structure. This process selects specific martensite twin variants in a biased fashion. The resulting as-trained material consists entirely of martensite (Fig. 6h). It exhibits finer lath-shaped martensite compared to that of the as-homogenized sample (*cf.*, Fig. 6g). Subsequently, the samples are thermally cycled up to 393 K (above $A_f$) to revert them to austenite and then cooled without any applied stress. As shown in Fig. 6i, the morphology of the samples after the thermal cycle is very similar to the as-trained case (*cf.*, Fig. 6h). The orientation match between the martensite structures of the as-homogenized and thermally treated samples is confirmed in Fig. S13. The experimental results are consistent with the above-discussed MD results (Figs. 3 and 4). The similarity between the experimental and theoretical microstructural evolution implies that the two-way shape-memory effect is the common origin. However, the primary source of the training-induced residual strain in the experiment is likely different than that in the MD simulations due to the significant grain size difference. In bulk SMAs with comparably large grains, the mechanical training process for the two-way shape-memory effect leads to an anisotropic dislocation structure within the austenitic matrix phase. This dislocation structure generates an anisotropic stress field, which guides martensite formation into preferentially oriented variants according to the deformation introduced during training, thereby causing macroscopic shape changes [45-47]. Furthermore, martensite reorientation during training in a polycrystalline matrix also effectively introduces internal plastic deformation [48, 49]. This internal plastic deformation is crucial for accommodating orientation mismatch among preferentially oriented martensite variants in adjacent grains.

The samples are subjected to further thermal cycles. After each thermal cycle, the shape recovery and thermal expansion during heating are measured to examine the effect of the thermomechanical training. Thermal expansion in each direction is measured on heating between 233 K and 313 K (Fig. 7c). These temperatures are below the $M_f$ temperature of 315 K. Therefore the thermal expansion is unaffected by the thermally-induced transformation. The measured CTE values and shape recovery are summarized in Figs. 7a and 7b, respectively. The as-homogenized sample exhibits an isotropic CTE value of $7.8 \pm 0.6 \times 10^{-6}$ K$^{-1}$, irrespective of the axis. The thermomechanical training clearly induces an anisotropic CTE. The positive thermal expansion along the compression direction (*Y*) is enhanced in the as-trained samples. In contrast, the CTE for the perpendicular directions (*X* and *Z*) becomes near zero (-0.043 – 0.91 × 10$^{-6}$ K$^{-1}$ for 5%-M3 case) or even negative (-5.3 – -2.3 for 5%-M10 and -4.4 – -1.1 × 10$^{-6}$ K$^{-1}$ for 10%-M1). For the as-trained state, the anisotropy of the CTE in 5%-M10 and 10%-M1 is more noticeable than in 5%-M3 ($\varepsilon_{5\%-M3} < \varepsilon_{5\%-M10} \approx \varepsilon_{10\%-M1}$). The anisotropy correlates with the degree of training, with higher strains and more mechanical cycles leading to a larger CTE anisotropy in the as-trained state, consistent with the MD simulation results (Fig. 5).

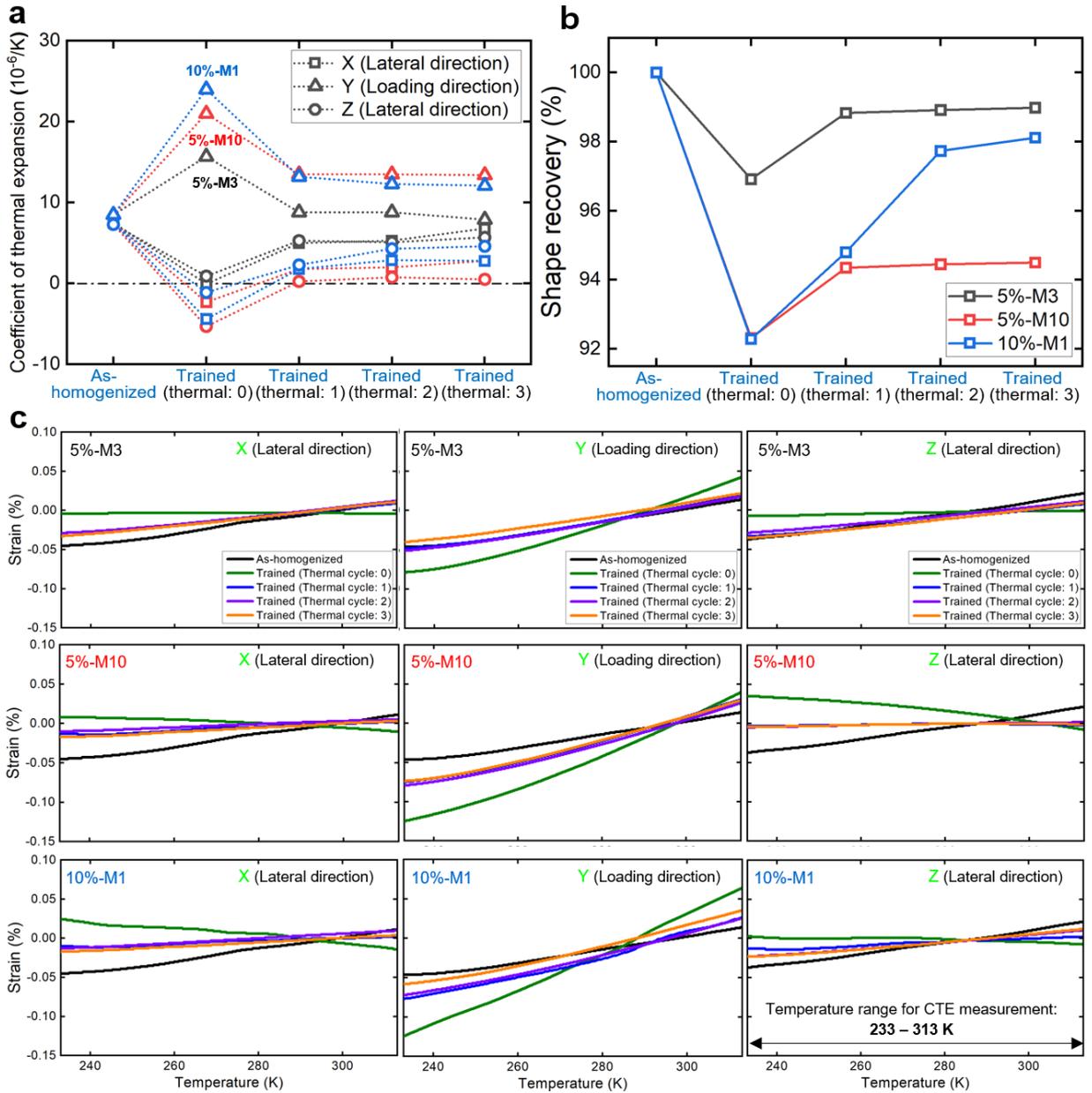

**Figure 7.** Experimental demonstration of tailoring the thermal expansion. (a) Measured coefficients of thermal expansion (CTE) along the axes as defined in Fig. 6a. (b) Shape change during thermomechanical training and thermal cycles (*N*). (c) Temperature dependence of the directional strain of as-homogenized and thermomechanically trained samples during heating (233 K → 313 K).

We consider in total three stress-free thermal cycles after the mechanical training (hereafter, labeled with an additional identifier for the number of thermal cycles, *e.g.*, 5%-M10-T3 for the case of ten mechanical cycles with a maximum strain of 5% and three subsequent thermal cycles). After thermal cycling, the biased CTE anisotropy is reduced, demonstrating that thermal training can be an additional processing condition for fine-tuning the CTE. This is again consistent with the MD results and, further, with the degradation of the two-way shape-memory effect reported by previous experiments on SMAs [50]. The most significant change in the CTE occurs after the first thermal cycle (*N*=1), and the variation in the 10%-M1-T(0 *vs.* 1) case is larger than in the 5%-M10-T(0 *vs.* 1). This trend corresponds to the shape recovery in Fig. 7b. The amount of shape recovery is limited and directly saturates during the first thermal cycle in the 5%-M3 and 5%-M10 cases. On the contrary, the shape recovery of the 10%-M1 case is comparably continuous with the number of thermal cycles. This means the cyclic deformation (5%-M10) is more effective in obtaining a stable biased twin variant selection than a single deformation (10%-M1). The training-induced anisotropic CTE in the 5%-M3-T(0-3) case gradually approaches the average CTE of a homogenized NiTi cube, and the average CTE along all directions becomes $6.8 \pm 1.1 \times 10^{-6}$ K$^{-1}$ at 5%-M3-T3. Interestingly, for the 5%-M10-T(1-3) case, the tailored CTE remains stable in a range of $0.26 - 1.8 \times 10^{-6}$ K$^{-1}$ (5%-M10-T1) and $0.52 - 2.8 \times 10^{-6}$ K$^{-1}$ (5%-M10-T3) even after the thermal cycling above the $A_f$ temperature. This behavior is analogous to previous experimental results on the two-way shape-memory effect, where the transformation strain induced by the mechanical training process was shown to saturate with an increased number of thermal cycles [50-52].

Overall, our results demonstrate the benefit of applying multiple mechanical training cycles to enhance the stability of the martensite microstructure. We expect that if the material is used at temperatures lower than the $A_f$ temperature, the gradual change in the CTE will not be a significant issue in environments with constantly changing temperatures because no phase transformation is involved. However, the enhanced microstructural stability achieved by cyclic mechanical training ensures reliable functionality, even under unexpected and high-temperature exposure above the $A_f$ temperature.

## 4. Summary and conclusion

Based on a combination of theoretical prediction and experimental verification, we have established an efficient procedure for tailoring the CTE of NiTi alloys. All the measured CTE values are summarized and compared to previous data in Fig. 1. The obtained minimum range of the in-plane CTE (-0.043 – 0.91 × $10^{-6}$ $K^{-1}$) is smaller than that of FeNi-based Invar alloys (~1.2 × $10^{-6}$ $K^{-1}$), and this is likely to be further reduced through finer adjustments of the thermomechanical process conditions in future studies, considering other compositions or even different material systems. Our proposed methodology demonstrates that by adjusting three key parameters—maximum load, mechanical cycle number, and thermal cycle number—the CTE of the material can be finely tuned to achieve the desired ZTE property precisely. Furthermore, the method suggested here is not specific to the ZTE but can also be used to fine-tune the CTE of materials in applications where the CTE of two materials must be matched, *e.g.*, composites.

The key asset of the present procedure is that the near ZTE along the in-plane directions is obtained by applying a simple and easily adjustable thermomechanical process based on a moderate level of mechanical loading, *i.e.*, cyclic mechanical thermal training. By avoiding severe plastic deformation and instead employing cyclic deformation, we achieve significantly reduced in-plane anisotropy in the CTE and pursue the use of the developed materials in a broader range of product forms not confined to bulk structural applications. Essential applications of materials with suppressed CTE are electronic packaging [22-24], shadow masks of display panels [25-27], and MEMs/NEMs [28-30].

The physical mechanism that underlies the presented CTE tailoring procedure is akin to the two-way shape-memory effect, as revealed by the MD simulations. Our analysis visually and directly validates the emergence of the two-way shape-memory effect and the resulting CTE properties by presenting step-by-step the microstructural changes during the training process. Accumulated plastic deformation due to training serves to memorize the orientation of nearby martensite during the initial mechanical treatment. The memory guides martensite formation in the following thermal treatment and induces anisotropy in the CTE. For nanocrystalline cells practically accessible with MD simulations, the high grain boundary area-to-volume ratio enhances the memory effect. For experimental polycrystalline samples with large grains, additional plastic deformation mechanisms likely act as a memory source. Previous experiments on the two-way shape-memory effect revealed [45, 51, 53] that irreversible strain can be induced by mechanical training of SMAs with comparably large grain sizes via dislocation slip. We thus expect that different sources of plastic deformation, *e.g.*, grain boundary sliding and dislocation slip, can help to memorize the specific martensite variants and to induce the anisotropy in the thermal expansion, with the detailed balance depending on factors such as grain size and dislocation density of the materials.

We expect the devised CTE tailoring procedure will expand the possible choices of Invar-like materials well beyond NiTi SMAs. There is a wide variety of reported SMA systems [54], and several SMA systems were already reported to exhibit a NTE/ZTE under certain conditions [13]. The method suggested here can help develop strategies for fine-tuning the CTE of those SMA systems for various applications that are not just confined to tailored thermal expansion. For example, there has been an increasing demand for multi-functional Invar-like materials that fulfill adjustable thermal expansion and other important properties (*e.g.*, high thermal conductivity and lightweight).

We suggest future research be directed towards raising the phase transformation temperatures of SMAs. Compared to FeNi-based Invar alloys that are limited to a specific temperature range for proper operation, a significant advantage of SMAs with tailored thermal expansion is their ability to maintain the NTE/ZTE over a broader range of temperatures as long as the martensitic structure is maintained. By increasing the phase transformation temperature (*i.e.*, the $A_f$ temperature), the operational temperature range can be thus extended.

**Acknowledgements**

The funding by the European Research Council (ERC) under the European Union's Horizon 2020 research and innovation programme (Grant Agreement No. 865855) is gratefully acknowledged. This work was also supported by the National Research Foundation of Korea (NRF) funded by the Ministry of Science and ICT (Grant No. RS-2024-00345442 and 2022M3H4A3095290), and the National Supercomputing Center with supercomputing resources including technical support (KSC-2022-CRE-0386). E.L.P. acknowledges financial support from the NSF Graduate Research Fellowship Program (Grant No. DGE-1745302).

**Data Availability**

The data that support the finding of this study is available from the corresponding author (email: wonsko@inha.ac.kr) upon reasonable request.

**Conflict of Interest**

The authors declare no conflict of interest.

**Author Contribution**

**Won Seok Choi:** Methodology, Investigation, Writing – original draft, Writing – review & editing. **Won-Seok Ko:** Conceptualization, Supervision, Methodology, Investigation, Writing – original draft, Writing – review & editing, Funding acquisition. **Yejun Park:** Methodology, Investigation, Writing – original draft, Writing – review & editing. **Edward L. Pang:** Methodology, Investigation, Writing – original draft. **Jong-Hoon Park:** Formal analysis, Visualization. **Hye-Hyun Ahn:** Formal analysis, Visualization. **Yuji Ikeda:** Methodology, Writing – review & editing. **Pyuck-Pa Choi:** Methodology, Writing- Reviewing and Editing. **Blazej Grabowski:** Supervision, Methodology, Writing – review & editing, Writing – original draft, Funding acquisition.

# Supplementary materials for

# "Finely tunable thermal expansion of NiTi by stress-induced martensitic transformation and thermomechanical training"

Won Seok Choi [a,b,1], Won-Seok Ko [c,1,*], Yejun Park [a,1], Edward L. Pang [b], Jong-Hoon Park [d], Hye-Hyun Ahn [d], Yuji Ikeda [e], Pyuck-Pa Choi [a], and Blazej Grabowski [e]

[a] Department of Materials Science and Engineering, Korea Advanced Institute of Science and Technology, Daejeon 34141, Republic of Korea.

[b] Department of Materials Science and Engineering, Massachusetts Institute of Technology, 77 Massachusetts Avenue, Cambridge, MA 02139, USA.

[c] Department of Materials Science and Engineering, Korea University, Seoul 02841, Republic of Korea.

[d] Department of Materials Science and Engineering, Inha University, Incheon 22212, Republic of Korea.

[e] Institute for Materials Science, University of Stuttgart, Pfaffenwaldring 55, Stuttgart 70569, Germany.

E-mail: wonsko@korea.ac.kr (W.-S. Ko)

**The file includes:**

**Supplementary Notes. S1-S5**

**Supplementary Tables. S1-S2**

**Supplementary Figures. S1-S13**

**Supplementary references**

**Supplementary Note S1.** **Reliability of the interatomic potential**

The interatomic potential [1] used in the present study is based on the second nearest neighbor modified embedded-atom method (2NN MEAM) formalism. The potential is optimized to capture temperature- and stress-induced phase transformations between the B2 austenite and B19′ martensite phases in equiatomic NiTi. We have performed additional simulations to validate the performance of the potential concerning the thermal expansion of polycrystalline NiTi, especially of the B19′ martensite structure. Fig. S2 shows the temperature dependence of the structural parameters, *i.e.*, the lattice constants and the monoclinic angle, predicted by the interatomic potential [1] in comparison with reported experimental data [2]. The systematic offset observed with respect to the experimental data originates from the density-functional-theory data used as input for the fitting and is a well-known phenomenon [3]. This offset does not affect the temperature dependence of the material properties. The results show a nearly constant $a$ lattice constant, an increase in the $b$ lattice constant, a decrease in the $c$ lattice constant, and a decrease in the monoclinic angle $\beta$ with increasing temperature. All temperature dependencies are in very good agreement with the experimental data near room temperature (black circles in Fig. S2).

The performance of the interatomic potential was further examined by comparing the temperature dependence of the interplanar spacing of the B19′ martensite structure, as listed in Table S2. Previous experiments [4-6] and DFT calculations [7] reported either a PTE or NTE of the B19′ martensite phase perpendicular to specific crystal planes. The prediction by the interatomic potential is consistent with the reported trend in the interplanar spacing, which again confirms the reliability of the potential concerning its thermal expansion behavior.

**Supplementary Note S2.** **Martensite microstructures formed in the MD simulations**

Figs. S3 and S4 show the stress-strain response and microstructural evolution of a nanocrystalline cell with an average grain size of 16 nm at 300 K under compressive and tensile loading, respectively. Especially, Figs. S3d and S4d visualize the local lattice orientation (LLO) of the nanocrystalline cell during the mechanical and thermal loading. After the austenite to martensite transformation, the color of each grain changes, reflecting the formation of the martensite phase as a structure of fine (001)B19′ compound twins. In some cases, thin regions of the austenite phase remain between the martensite phase at maximum loading (marked in Figs. S3c and S4c by "Domain boundaries"). These regions correspond to martensitic domain boundaries, which divide multiple martensite domains. In some cases, this twinned structure occurs with multiple domains divided by domain boundaries. When martensite nucleation begins simultaneously at various sites of the grains, there should be an orientation mismatch between regions originating from different martensite nuclei. If this orientation mismatch is not resolved during the process of the martensite growth, a residual austenite (or domain boundary) can be present in the grain. A similar, twinned microstructure was also observed in previous experiments on the phase transformation of NiTi SMAs at the nanoscale, and it was referred to as a "herringbone" structure [8]. Finely dispersed (001)B19′ compound twin boundaries and the herringbone structure were also reported in previous MD studies based on the same interatomic potential used in the present study [9].

**Supplementary Note S3.** **Impact of loading and heating/cooling rate in the MD simulations**

We examined the impact of loading rate on the thermomechanical training process in the MD simulations. As shown in Fig. S9a, an increased stress rate weakens the mechanical training effect (*i.e.*, the two-way shape-memory effect decreases). This behavior is a consequence of a decreasing maximum training strain for higher stress rates. As shown in Fig. S10, an increased stress rate results in an overall increase in the slope of the stress-strain curves and the disappearance of the plateau stress. Such a behavior was also reported in a previous MD study [9]. Because the mechanical loading in the present MD runs was performed by stress control, the maximum reachable strain at constant maximum stress decreases with increasing stress rate.

Previous experimental studies [10, 11] involving compressive loading and unloading at temperatures above $A_\mathrm{f}$ show a similar behavior as the MD results, *i.e.*, an overall increase in the slope of the stress-strain curves and the disappearance of the plateau stress with increasing loading rate. This suggests that higher strain rates may result in less uniform martensite growth, potentially causing instability in the two-way shape-memory effect and the ZTE/NTE properties during thermal cycling.

Concerning thermal cycling, the MD simulations show that an increase in cooling/heating rates results in an overall weakening of the thermal training effect, as exemplified in Fig. S9b. At lower cooling/heating rates, sufficient relaxation time allows for the reorganization of martensite variants and stress perturbation damping, resulting in a more pronounced decrease in the CTE anisotropy.

**Supplementary Note S4.** **CTE calculations based on the GNLPTM**

Transformation twinning modes in NiTi depend on processing conditions and have been categorized as follows: Type I/II twins known from conventional martensitic transformations of bulk materials, a herringbone arrangement of (001)B19′ compound twins, and others [8, 12-17]. It has been reported that solutionized NiTi most frequently exhibits Type II twins [14]. An intermediate R phase can be induced by cold working, aging, and the addition of a third element (*e.g.*, Fe), and it is related to the formation of (001)B19′ twin compounds [18-22]. Finely dispersed (001)B19′ twin compounds were usually observed in nanocrystalline NiTi in experiments[8] and MD simulations [1, 9], including the present work (Note S2).

We performed theoretical calculations based on the phenomenological geometrically nonlinear theory of martensite (GNLPTM) to examine the effect of different twinning modes on the mechanically induced anisotropic CTE of martensite. Such calculations are complicated because martensite formed during the thermally-induced or stress-induced transformation is twinned to satisfy the crystallographic theory of martensite [23-28]. Prior work only considered the anisotropy of the martensite B19′ CTE tensor, which implies a single correspondence variant; this is sufficient for studies in which martensite is thermomechanically processed in the martensite phase to promote detwinning and variant conversion towards a single variant. Here, we have not performed any additional processing after transformation to disrupt the twinned martensite microstructure, and thus, the CTE calculations must consider the fractions and orientations of twins.

Predicted CTE values for stress-induced martensite in uniaxial compression and tension are given as a function of austenite orientation in Fig. S12 for untwinned (single correspondence variant), (011) Type I, (111) Type I, [011] Type II, [211] Type II, and (001) compound twinning modes. The [011] Type II twinning mode is widely accepted as the most favorable in large-grained solutionized NiTi for both thermally- and stress-induced martensite [14, 21, 29-31]. The results show a clear variation in the CTE of the stress-induced martensite as a function of the austenite orientation. Discontinuities in the CTE can also be seen when the preferred habit plane variant (HPV) changes.

The CTE along the loading direction of each stress-induced martensite is averaged over all austenite orientations and listed in Fig. S12. Comparing different twining modes, while the absolute values of the averaged CTE change, the signs of them are consistent. Interestingly, the trends observed by the present MD simulations are consistently present regardless of the type of twinning mode, *i.e.*, PTE for the compressive direction and NTE for the tensile direction.

**Supplementary Note S5. Uncertainty in the CTE measurement**

We present a quantitative assessment of the uncertainty in the CTE measurements. Typically, the data obtained from the thermomechanical analyzer (TMA) (*e.g.*, Fig. 7c) are fitted to a straight line using the least squares method, and the CTE value is obtained from the slope of this line. The CTE value is expressed as

$$CTE = \frac{n\sum_{i=1}^{n}\varepsilon_i T_i - \sum_{i=1}^{n}\varepsilon_i \sum_{i=1}^{n} T_i}{n\sum_{i=1}^{n} T_i^2 - \left(\sum_{i=1}^{n} T_i\right)^2}, \quad (1)$$

where $T_i$ and $\varepsilon_i$ are the temperature and strain for the *i*'th data point, and $n$ is the number of data points. The standard deviation of the CTE ($\sigma_{CTE}$) is described by:

$$\sigma_{CTE} = \sqrt{\frac{SSE}{(n-2)\sum_{i=1}^{n}(T_i - \overline{T})^2}}. \quad (2)$$

When accounting for the inherent uncertainty of the TMA equipment, the sum of the squared error (SSE) [32] is given by

$$SSE \approx a^2 \cdot (n-2), \quad (3)$$

where $a$ is the inherent uncertainty strain (*i.e.*, $a$ = 0.125 nm / 6 mm). The combination of Eqs. (2) and (3) yields the following equation:

$$\sigma_{CTE} = \frac{a}{\sqrt{\sum_{i=1}^{n}(T_i - \overline{T})^2}}. \quad (4)$$

In the present study, the TMA measurement was conducted at 0.1 K intervals over a temperature range of 233 K to 313 K. Applying this range to Eq. (4), the standard deviation of the CTE obtained from the TMA is $3.18 \times 10^{-11}$. Therefore, it can be concluded that the uncertainty in the CTE measurements presented in this study is marginal compared to the differences in the measured CTE for each condition.

## Supplementary tables

**Table S1.** Detailed conditions considered in the present MD simulations. Average grain diameter ($D$), number of grains, cell dimensions, and number of atoms in the nanocrystalline cells are presented.

| $D$ (nm) | 8 | 10 | 12 | 14 | 16 |
|---|---|---|---|---|---|
| Number of grains | 30 | 30 | 30 | 30 | 30 |
| Cell dimensions (nm) | 20×20×20 | 25×25×25 | 30×30×30 | 35×35×35 | 40×40×40 |
| Number of atoms | $5.93 \times 10^5$ | $1.16 \times 10^6$ | $2.02 \times 10^6$ | $3.21 \times 10^6$ | $4.78 \times 10^6$ |

**Table S2.** Thermal expansion behavior of the B19′ martensite structure perpendicular to specific crystal planes, summarized as positive thermal expansion (PTE) and negative thermal expansion (NTE). The prediction by the 2NN MEAM potential [1] used for the present MD simulations is compared with the previous experiments by Qiu *et al*. [4] and Ahadi *et al*. [5, 6], DFT calculations based on the quasiharmonic approximation (QHA) by Ahadi *et al*. [6], and *ab initio* MD by Haskins and Lawson [33].

| Plane | Experiment [Qiu][4] | Experiment [Ahadi][5, 6] | DFT (QHA)[6] | DFT (MD)[33] | 2NN MEAM (MD)[1] |
|---|---|---|---|---|---|
| $(10\bar{2})$ | | *NTE* | *NTE* | *NTE* | *NTE* |
| $(11\bar{2})$ | | *NTE* | *NTE* | *NTE* | *NTE* |
| $(002)$ | | *NTE* | *NTE* | *NTE* | *NTE* |
| $(10\bar{1})$ | | *NTE* | *NTE* | *NTE* | *NTE* |
| $(12\bar{2})$ | | *NTE* | *NTE* | *NTE* | *NTE* |
| $(11\bar{1})$ | | *NTE* | *NTE* | *NTE* | *NTE* |
| $(011)$ | | PTE | PTE | PTE | PTE |
| $(12\bar{1})$ | | PTE | *NTE* | PTE | PTE |
| $(021)$ | | PTE | PTE | PTE | PTE |
| $(103)$ | | PTE | PTE | PTE | PTE |
| $(020)$ | PTE | PTE | PTE | PTE | PTE |
| $(120)$ | PTE | PTE | PTE | PTE | PTE |
| $(100)$ | | PTE | PTE | PTE | PTE |
| $(112)$ | | PTE | PTE | PTE | PTE |
| $(102)$ | | PTE | PTE | PTE | PTE |
| $(111)$ | PTE | | | | PTE |
| $(001)$ | PTE | | | | *NTE* |
| $(110)$ | PTE | | | | PTE |
| $(10\bar{1})$ | *NTE* | | | | *NTE* |
| $(20\bar{1})$ | *NTE* | | | | *NTE* |

## Supplementary figures

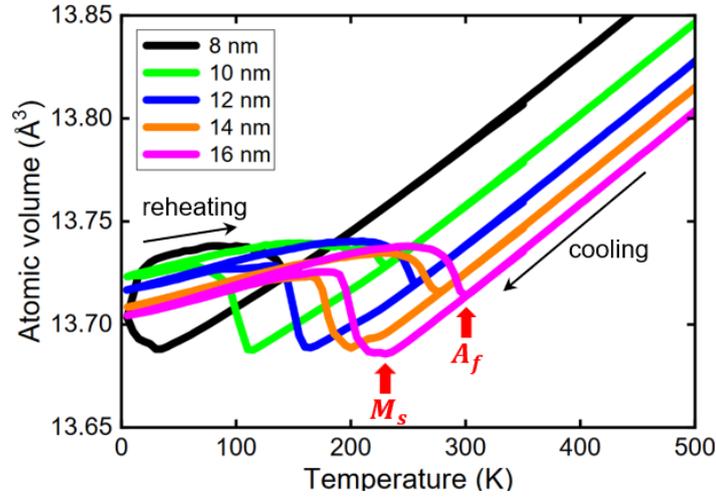

**Figure S1.** Temperature dependence of the atomic volume in a nanocrystalline cell with various average grain sizes under no external loading, as predicted by the present MD simulations. The location of the martensite start ($M_S$) and austenite finish ($A_f$) temperature is indicated by the red arrows for the 16 nm curve.

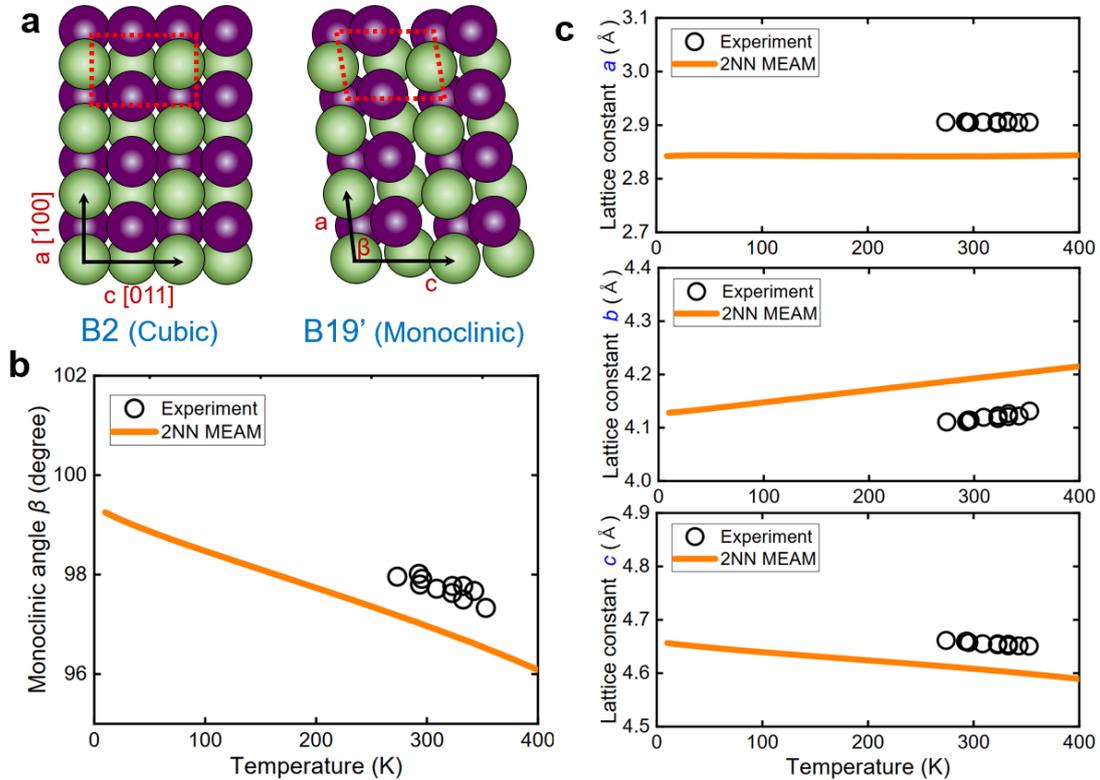

**Figure S2.** Temperature dependence of the structural parameters. (a) Atomic structures of the cubic B2 (austenite) and monoclinic B19′ (martensite) phases with the monoclinic angle ($\beta$) and the lattice constants ($a$ and $c$) indicated. Ni and Ti atoms are represented by purple and green balls, respectively. Temperature dependence of (b) the monoclinic angle ($\beta$) and (c) lattice constants ($a$, $b$, and $c$) of the B19′ of the B19′ martensite structure calculated using the 2NN MEAM potential [1] used for the present MD simulations, in comparison with experimental data [2].

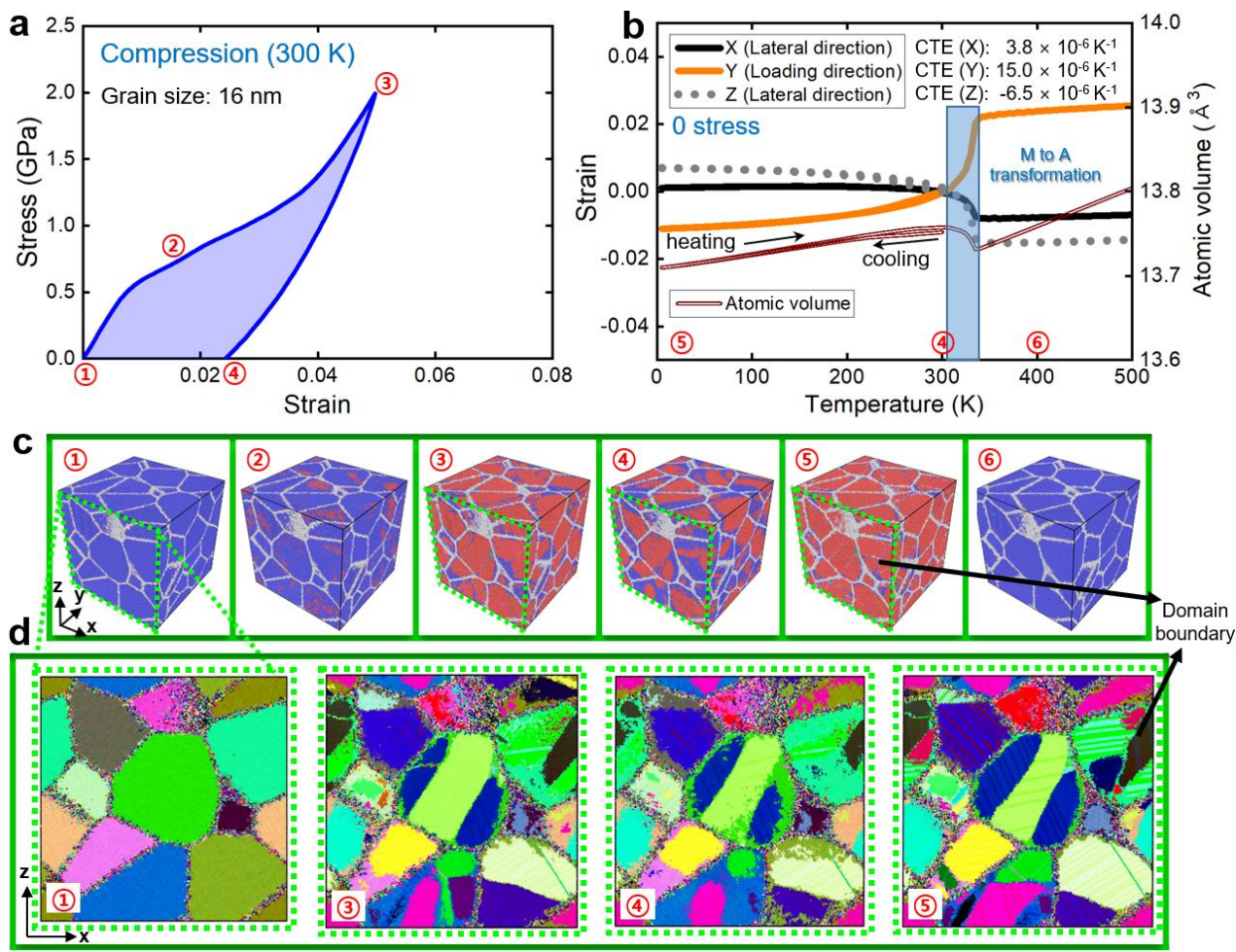

**Figure S3.** Impact of compressive loading on the thermal expansion behavior and microstructural evolution, predicted by the present MD simulations. (a) Stress-strain response of a nanocrystalline cell with an average grain size of 16 nm under mechanical (compressive) loading and unloading at 300 K. (b) Temperature dependence of the directional strain and atomic volume of the nanocrystalline cell during the subsequent thermal loading (300 → 5 → 500 K) at zero external stress. The reference of the strain values is the cell at 300 K. The coefficient of linear thermal expansion (CTE) measured in the temperature range of 5 – 100 K is indicated for each direction. (c) Corresponding atomic configurations visualized by the PTM algorithm [34]. In each snapshot, blue atoms correspond to the B2 austenite structure, red atoms to the B19′ martensite structure, and gray atoms to the grain and domain boundaries. (d) Evolution of the microstructure during the mechanical and thermal loading, identified by the local lattice orientation (LLO) of the crystalline regions based on the PTM algorithm [34]. The color of the atoms corresponds to specific crystal orientations.

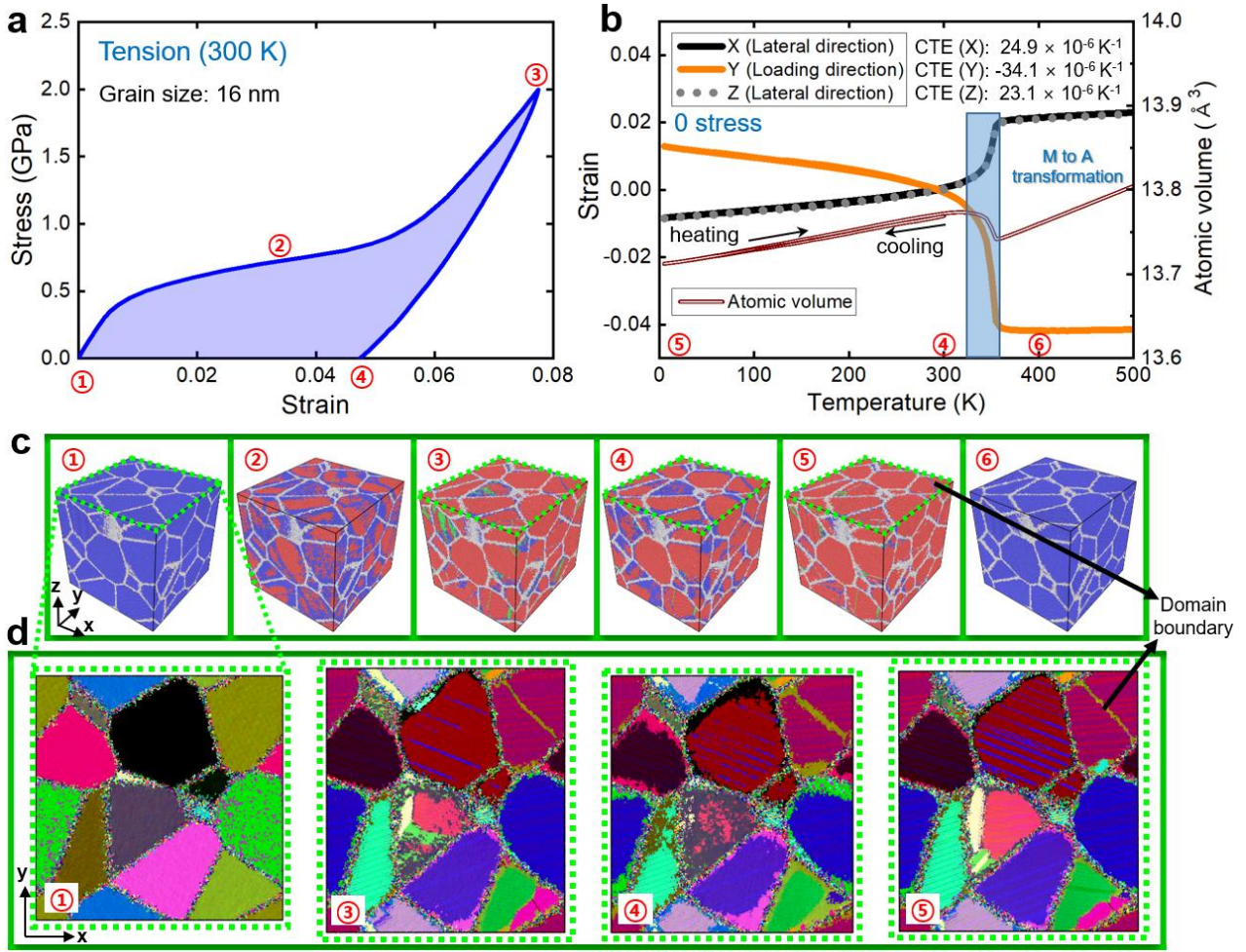

**Figure S4.** Impact of tensile loading on the thermal expansion behavior and microstructural evolution, predicted by the present MD simulations. (a) Stress-strain response of a nanocrystalline cell with an average grain size of 16 nm under mechanical (tensile) loading and unloading at 300 K. (b) Temperature dependence of the directional strain and atomic volume of the nanocrystalline cell during the subsequent thermal loading (300 → 5 → 500 K) at zero external stress. The reference of the strain values is the cell at 300 K. The coefficient of linear thermal expansion (CTE) measured in the temperature range of 5 – 100 K is indicated for each direction. (c) Corresponding atomic configurations visualized by the PTM algorithm [34]. In each snapshot, blue atoms correspond to the B2 austenite structure, red atoms to the B19′ martensite structure, and gray atoms to the grain and domain boundaries. (d) Evolution of the microstructure during the mechanical and thermal loading, identified by the local lattice orientation (LLO) of the crystalline regions based on the PTM algorithm [34]. The color of the atoms corresponds to specific crystal orientations.

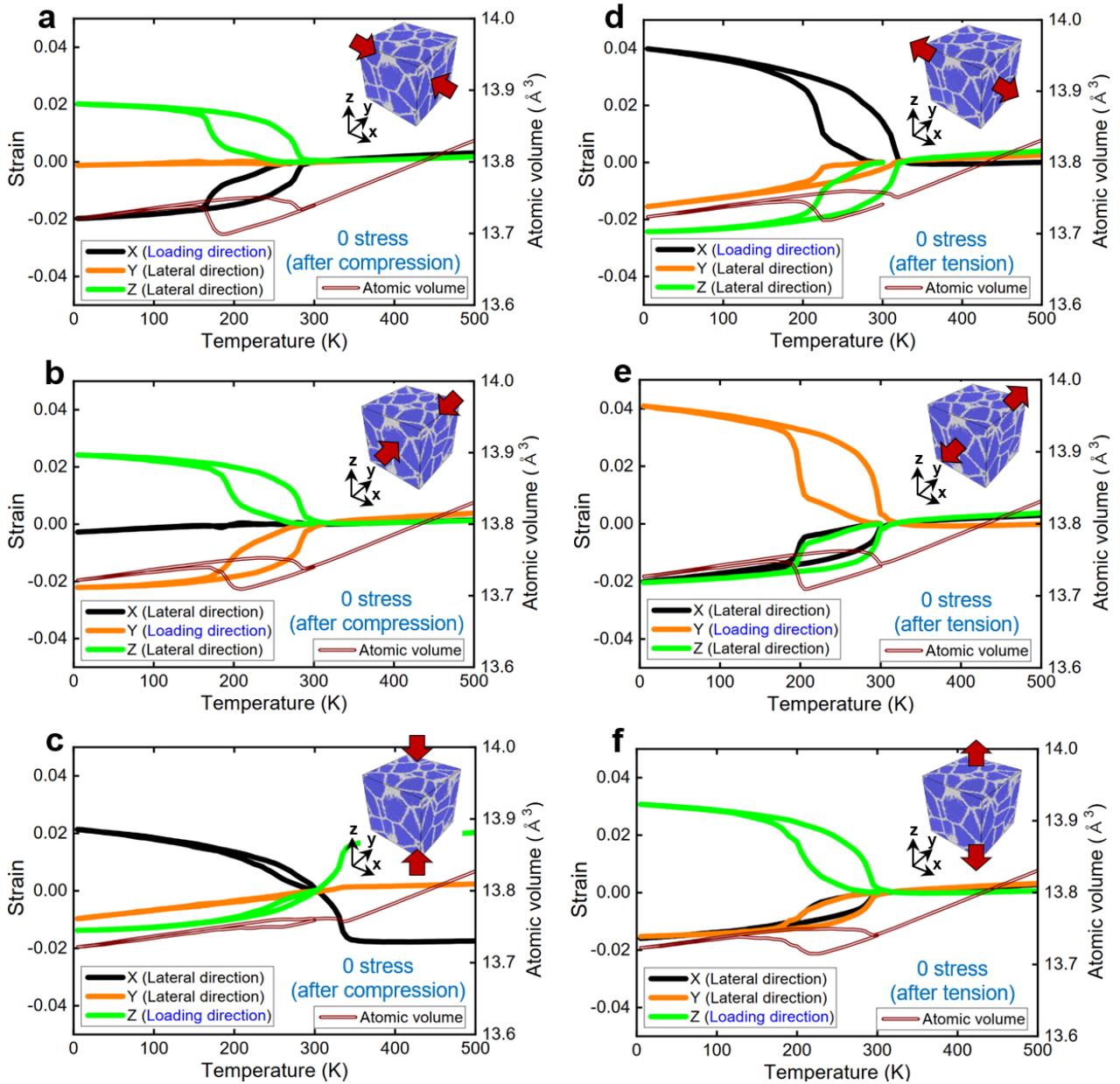

**Figure S5.** Temperature dependence of the directional strain and atomic volume of a nanocrystalline cell with an average grain size of 12 nm, predicted by the present MD simulations. Results of the nanocrystalline cell prepared by preceding (a–c) compressive and (d–f) tensile loading are presented. Results of the loading along (a,d) *X*-direction, (b,e) *Y*-direction, and (c,f) *Z*-direction are presented. The reference of the strain values is the cell at 300 K.

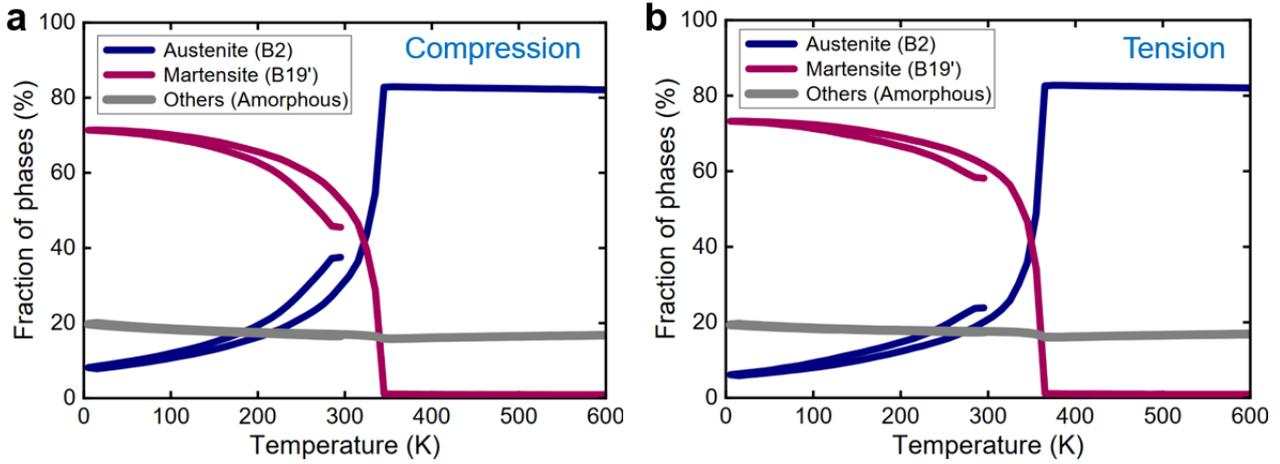

**Figure S6.** Temperature dependence of the phase fractions in a nanocrystalline cell with an average grain size of 16 nm, predicted by the present MD simulations. Results of the nanocrystalline cell prepared by preceding (a) compressive and (b) tensile loading are presented. The fractions of the B2 austenite, B19′ martensite, and other (mostly amorphous) structures are identified based on the PTM algorithm [34].

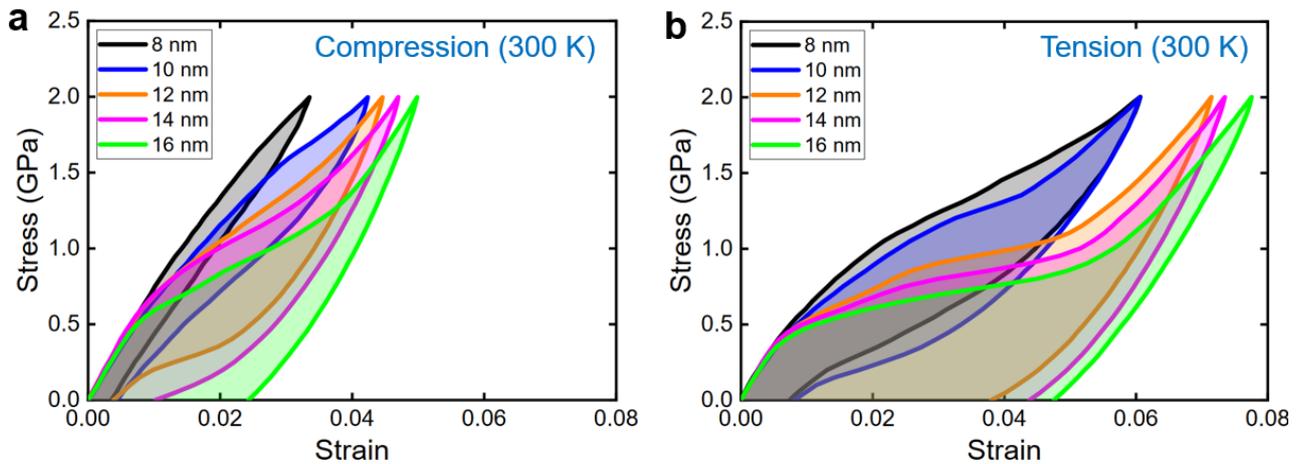

**Figure S7.** Stress-strain responses of nanocrystalline cells with different average grain diameters (8, 10, 12, 14, and 16 nm), predicted by the present MD simulations. Results of the nanocrystalline cells under a) compressive and b) tensile loading at 300 K are presented.

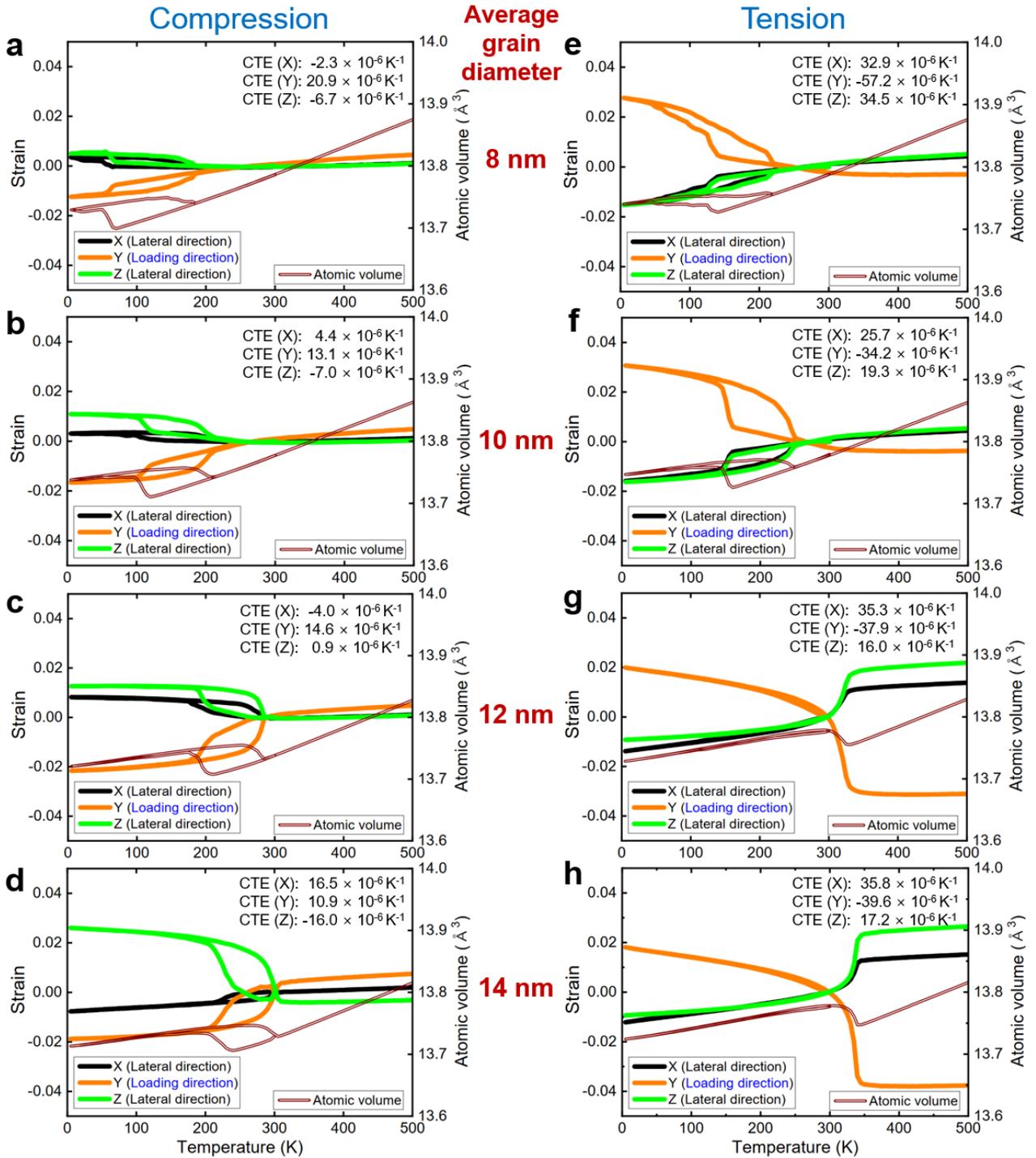

**Figure S8.** Temperature dependence of the directional strain and atomic volume of nanocrystalline cells with different average grain diameters (8, 10, 12, and 14 nm), predicted by the present MD simulations. Results of the nanocrystalline cells prepared by preceding (a–d) compressive and (e–h) tensile loading (Fig. S7) are presented. The reference of the strain values is the cell at 300 K.

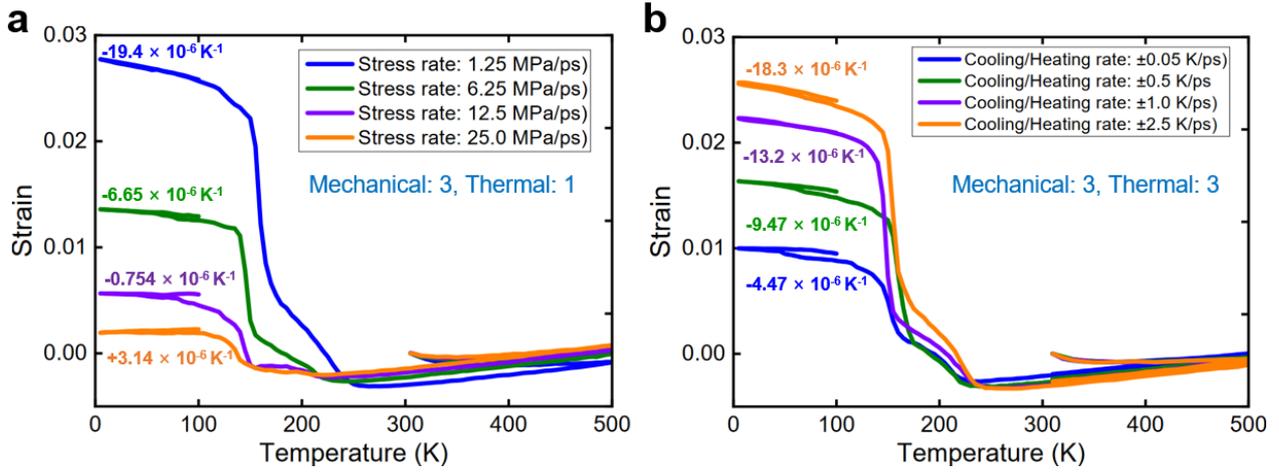

**Figure S9.** Temperature dependence of the directional strain in a nanocrystalline cell with an average grain diameter of 12 nm prepared by preceding mechanical (Fig. S10) and thermal cyclic loadings, predicted by the present MD simulations. The reference of the strain values is the cell at 300 K. (a) Impact of different strain rates for the mechanical loading. (b) Impact of different cooling/heating rates for the thermal loading.

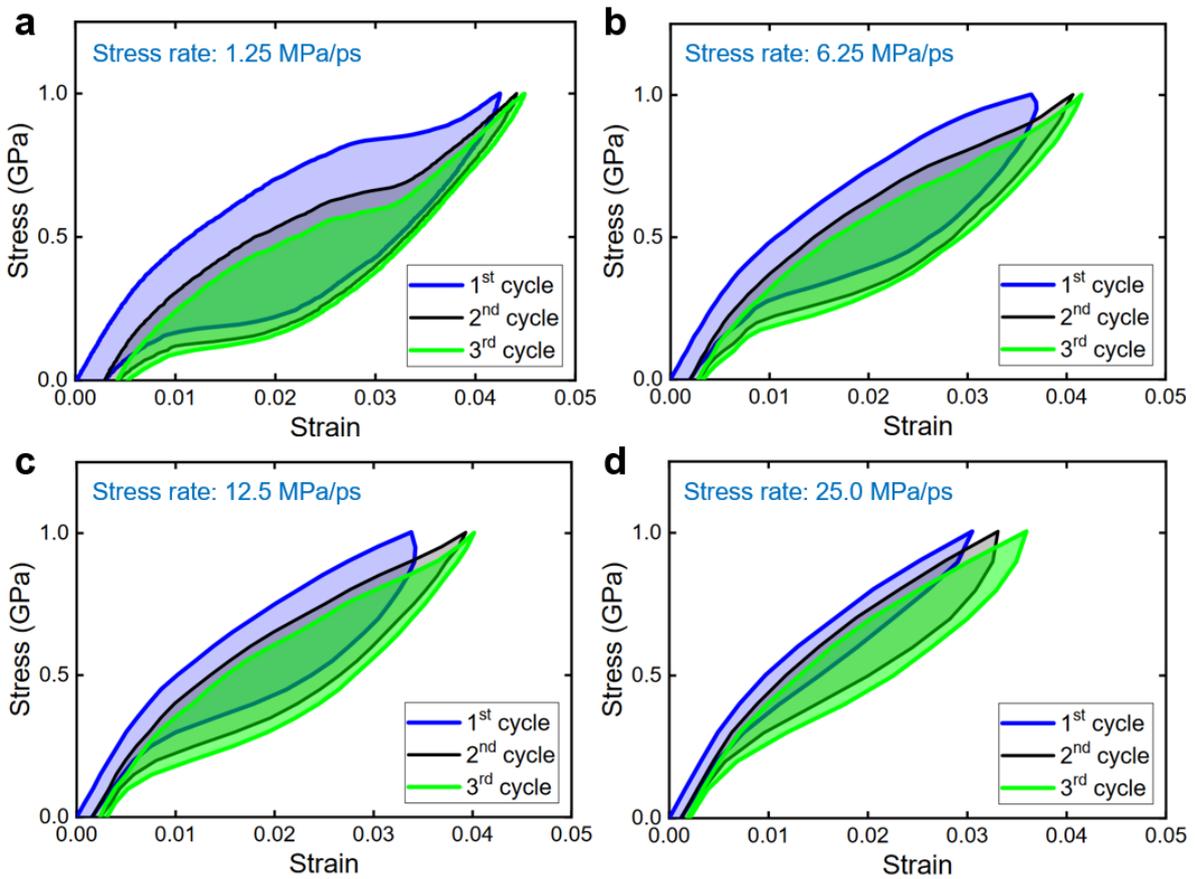

**Figure S10.** Impact of different strain rates on the mechanical behavior, predicted by the present MD simulations. (a–d) Stress-strain responses of a nanocrystalline cell with an average grain diameter of 12 nm under cyclic tensile loading at 300 K with strain rates of (a) 1.25, (b) 6.25, (c) 12.5, and (d) 25.0 MPa/ps are presented.

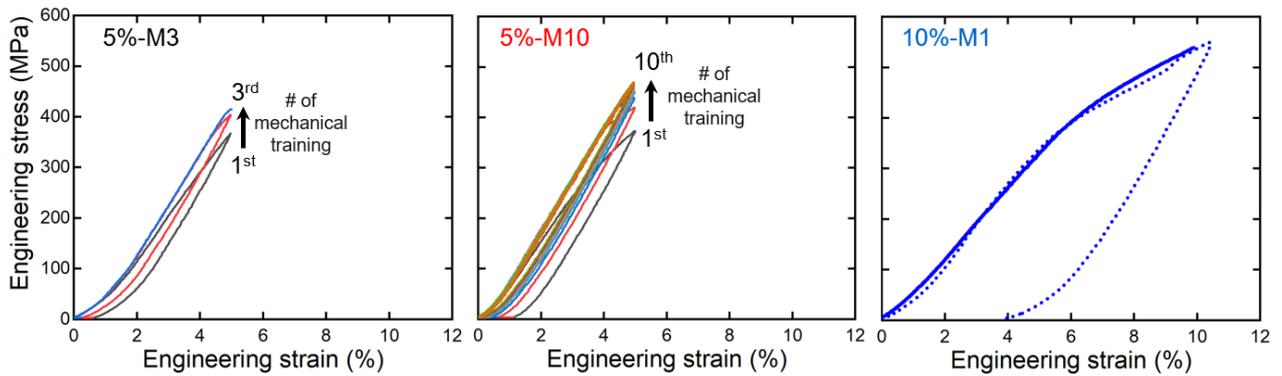

**Figure S11.** Engineering stress-strain curves of NiTi cube under mechanical training. Results of (a) 5%-M3, (b) 5%-M10, and (c) 10%-M1 for training at 120 °C are presented. The accumulated residual strain during training was measured to be (a) 0.9% and (b) 3.4%. (c) The residual strain during one-off deformation in 10%-M1 was measured from an independent compression test (dotted line) to be 3.9%. The applied stress increased to reach the target strain, and the samples were cooled down to room temperature with an applied stress of (a) 415 MPa, (b) 468 MPa, and (c) 539 MPa.

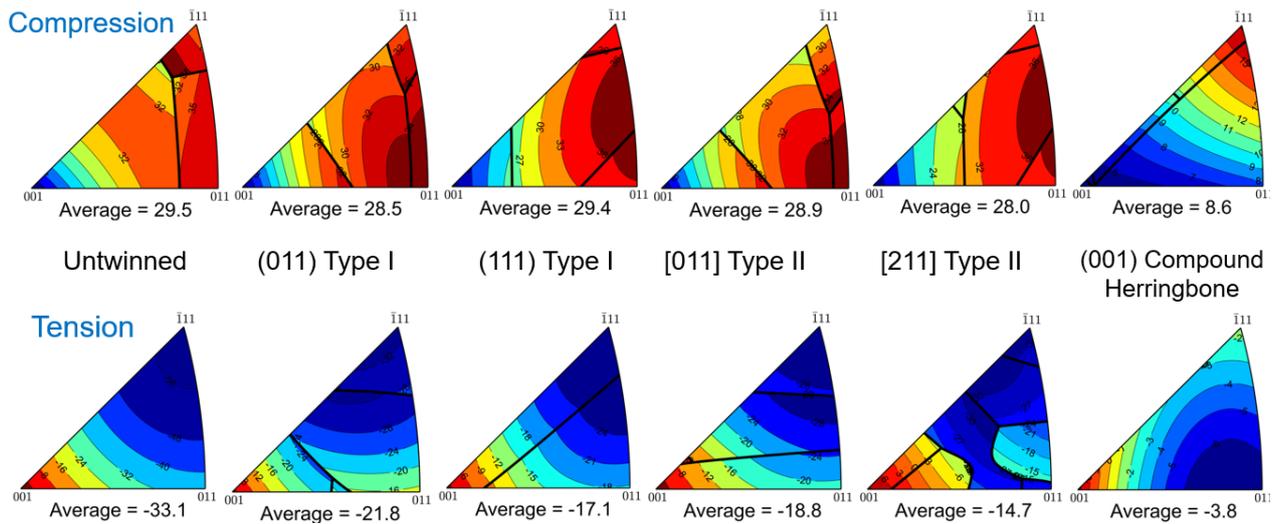

**Figure S12.** Calculated orientation dependence of the CTE in the loading direction for a stress-induced transformation in uniaxial compression and tension. Results are shown for the untwinned (single correspondence variant) case, all conventional twinning modes allowed by the crystallographic theory of martensite, and (001) compound twinning in a herringbone microstructure. Inverse pole figures are shown as a function of austenite orientation. Solid lines represent boundaries across which different martensite variants are selected. Average CTE values over all orientations are given below each subfigure. CTE values are given in $10^{-6}$ K$^{-1}$.

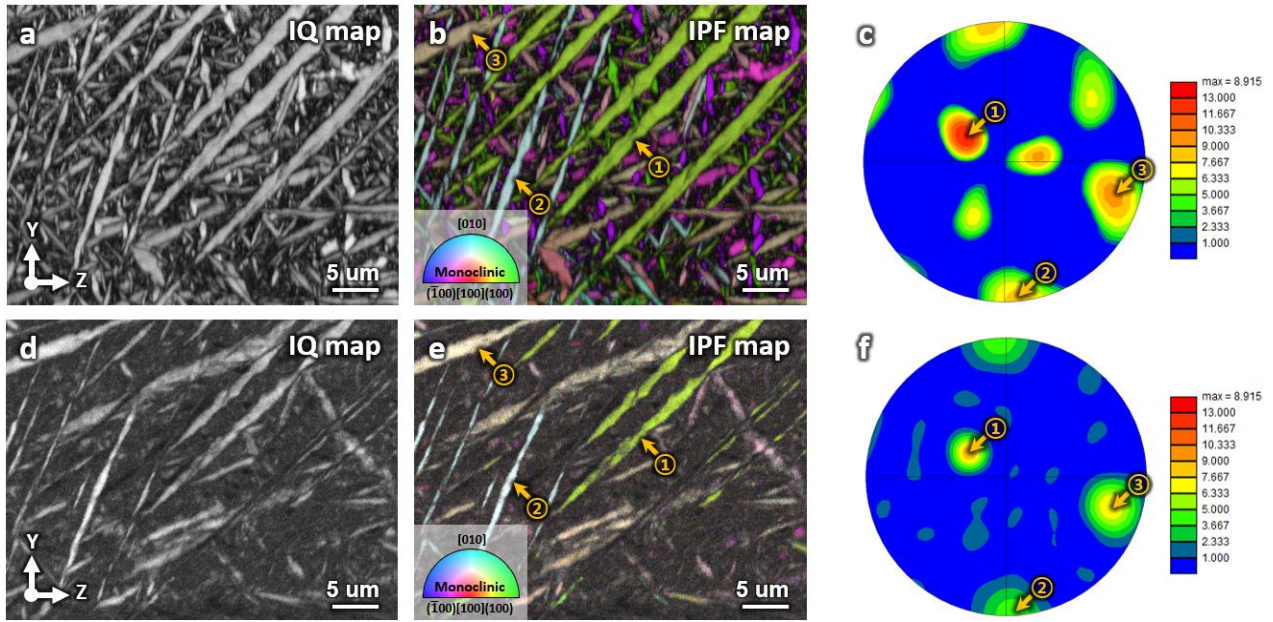

**Figure S13.** Experimentally analyzed results of NiTi cube sample. Image quality (IQ) map, inverse pole figure (IPF) map, and inverse pole figure of the specimen (a–c) after mechanical training (5%-M10) and (d–f) after additional thermal loading.

## Supplementary References